\documentclass[aps, prb, showpacs, twocolumn, 
amssymb,superscriptaddress]{revtex4}
\usepackage{bm, amsmath, amsfonts}
\usepackage[dvipdfmx]{graphicx}
\usepackage{float}
\usepackage{color}

\def\dag{\dagger}

\newcommand{\ii}{\text{i}}

\begin{document}


\title{
Exact zero modes in twisted Kitaev chains
}

\author{Kohei Kawabata}
\affiliation{Department of Physics, Graduate School of Science, The University of Tokyo, Hongo, Tokyo 113-0033, Japan}

\author{Ryohei Kobayashi}
\affiliation{Department of Physics, Graduate School of Science, The University of Tokyo, Hongo, Tokyo 113-0033, Japan}

\author{Ning Wu}
\affiliation{Center for Quantum Technology Research, School of Physics, Beijing Institute of Technology, Beijing 100081, China}

\author{Hosho Katsura}
\affiliation{Department of Physics, Graduate School of Science, The University of Tokyo, Hongo, Tokyo 113-0033, Japan}

\date{\today}

\begin{abstract}
We study the Kitaev chain under generalized twisted boundary conditions, for which both the amplitudes and the phases of the boundary couplings can be tuned at will. 
We explicitly show the presence of 
exact zero modes for large chains belonging to the topological phase in the most general case, in spite of the absence of ``edges" in the system. For specific values of the phase parameters, we rigorously obtain the condition for the presence of the 
exact zero modes in finite chains, and show that the zero modes obtained are indeed localized. The full spectrum of the twisted chains with zero chemical potential is analytically presented. Finally, we demonstrate the persistence of zero modes (level crossing) even in the presence of disorder or 
interactions. 
\end{abstract}


\maketitle

\section{Introduction}
\label{sec:intro}

Majorana zero modes have played an important role in condensed matter physics in recent years~\cite{Wilczek09,Alicea12,LeijnseFlensberg12,ElliottFranz15}. The Majorana modes are the same as their own antimodes by definition, and they have been anticipated to appear as zero-energy bound states. There have been considerable efforts towards the realization of Majorana modes in condensed matter settings~\cite{Mourik-12, Rokhinson-12, Das-12, DengYu-12, Finck-13, Churchill-13, Nadj-Perge-14, Higginbotham-15}. The existence of Majorana zero modes is of special interest because it can be applied to the physical construction of qubits for topological quantum computing~\cite{Ivanov-01, Nayak-08, Pachos-12}. From an experimental point of view, it is essential to investigate the effects of disorder~\cite{Brouwer-11, Lobos-12, DeGottardi-14, Altland-14, Crepin-14} and interactions~\cite{FidkowskiKitaev10, FidkowskiKitaev11, Turner11, Gurarie11, Gangadharaiah-11, Stoudenmire-11, Hassler-12, Thomale-13, Katsura_Int_Majorana, Rahmani-15-L, Rahmani-15-B, Gergs-16}. Furthermore, various theoretical aspects have been revealed, including the connection with supersymmetry~\cite{Grover-14, Jian-15, Rahmani-15-L, Rahmani-15-B, Hsieh-16}, the generalization to parafermion modes~\cite{Fendley12, Clarke-14, KlinovajaLoss14, Mong-14, Jermyn-14, Aris-15, Alicea-16, Iemini-16}, and the construction of topologically invariant defects~\cite{Aasen-16}.

The emergence of Majorana zero modes in condensed matter systems  
was first proposed by Kitaev~\cite{Kitaev01}. The Kitaev chain is a one-dimensional lattice model that describes a spin-polarized $p$-wave superconductor with open boundaries. This model possesses a topological phase with two-fold degenerate ground states which are robust against local perturbations that preserve the fermion parity symmetry. 
The origin of the ground-state degeneracy in the topological phase is the presence of Majorana zero modes. These zero-energy modes, often called strong zero modes,  commute with the Hamiltonian and anti-commute with the fermionic parity, and are localized near the boundaries. However, the existence of a topological phase can be manifested by only a minimal degeneracy, the ground-state degeneracy, say, leading to the notion of a `weak zero mode~\cite{Alicea-16}' that only commutes with the projected Hamiltonian onto the low-energy manifold. 
Moreover, the celebrated bulk-boundary correspondence is one of the most crucial properties for the topological phases of matter as a general rule, which topological insulators also exhibit~\cite{HasanKane10, QiZhang11, shortcourse topo.ins.}. Here, the fundamental question that should be addressed is whether Majorana zero modes (either strong or weak) can persist for a system without boundaries. Naively, one might think that there will be no zero-mode since there is no ``edge'' in the system.

In this paper, we will answer the above question by investigating the Kitaev chains with generalized twisted boundary conditions (TBCs), i.e., we arbitrarily control the amplitudes and the phases of the couplings on the boundaries. The usual periodic boundary condition (PBC) and the anti-periodic boundary condition (APBC) are included in the TBC as limiting cases. In practice, such a boundary condition can be realized by magnetic fluxes and Josephson junctions~\cite{Kitaev01, Kwon04, Fu-09, Jiang-11, San-Jose12, Rokhinson-12, Deng-12, Deng-13, Beenakker-13-Josephson, Hansen-16, Marra-16, Alase-16, Dmytruk-16, Giuliano-16, Cobanera-16}. Our work is motivated by the observation that the fermionic parities in the ground states of the Kitaev chain with the PBC and the APBC have opposite signs in the topological phase, hence indicating a level crossing when one continuously changes the parameters so as to connect the PBC with the APBC. When the open Kitaev chain resides in the topological phase, we show that for sufficiently large chains, Majorana zero modes do appear when the phase parameters at the boundary are tuned to 
specific values, in spite of the absence of the edges~\cite{def_Majorana_zero}. 
Then, we obtain the condition for the presence of Majorana zero modes in finite chains, as well as explicitly determine the spatial profile of the zero modes. 
We note that the emergence of Majorana zero modes (protected level crossing) in the fractional Josephson effect has been well-established since the seminal work of Kitaev~\cite{Kitaev01, Jiang-11, Alicea12, Alase-16, Dmytruk-16, Giuliano-16}. However, we emphasize that the boundary couplings studied in this paper are more general than those in previous work, as their amplitudes and phases can be arbitrary. More importantly, our derivation of the condition for the Majorana zero modes does not require the assumption that the Majorana edge modes that appear in the absence of the boundary term do not hybridize much with the bulk states, which was implicitly assumed from the outset in most previous work.

We also investigate the robustness of the zero modes against disorder and interactions. We show for a particular set of parameters that the Majorana zero operators that commute with the Hamiltonian exist even in the presence of spatially varying couplings. The level crossing signaling a topological order is found to be robust against nearest-neighbor 
interactions, given that the bulk parameters are those of the interacting Kitaev chain in the topological phase. This means that the zero modes (at least in the weak sense~\cite{Alicea-16}) survive even in the presence of interactions. 

The paper is organized as follows. In Sec.~\ref{sec:model}, we introduce the model, and point out that there should be Majorana zero modes when we smoothly change the boundary conditions between the PBC and the APBC. 
In Sec.~\ref{sec:emergence of zero modes}, we directly compute the Pfaffian of the Hamiltonian and show the precise condition for the existence of zero modes. 
In Sec.~\ref{sec:exact zero modes}, we present explicit forms of the zero modes and some other properties, including the full spectrum of the chain at zero chemical potential. 
In Sec.~\ref{sec: interaction}, we investigate the effects of nearest-neighbor 
interactions. We numerically demonstrate the presence of the zero modes in the weak sense 
and determine the condition for the emergence of the zero modes. 
We conclude this paper in Sec.~\ref{sec:conc}. In Appendix~\ref{APP_PAP}, we calculate in detail the fermionic parity of the ground state of the Kitaev chain for the PBC and the APBC. In Appendix~\ref{sec:zero mode matrix}, we present an explicit expression for the matrix which in general determines whether zero modes exist or not. In Appendix~\ref{sec:mu0 detail}, we give a detailed procedure for calculating all the eigen-energies of the chain with zero chemical potential. In Appendix~\ref{sec:frustration-free}, the presence of the zero modes in interacting chains is confirmed for a solvable (frustration-free) model\cite{Katsura_Int_Majorana}.

\section{Model and Phases}
\label{sec:model}

We consider a system of spinless fermions on a chain of length $L$. For each site $j=1,2, \cdots, L$, we denote by $c^{\dagger}_{j}$ and $c_{j}$ the creation and the annihilation operators, respectively. We impose a \textit{twisted boundary condition}, in which the amplitudes and the phases of the parameters on the boundaries can be tuned arbitrarily.

The Hamiltonian in question consists of two parts: $H = H_{\mathrm{bulk}} + H_{\mathrm{boundary}}$. The Hamiltonian for the bulk, $H_{\mathrm{bulk}}$, is described by
\begin{align}
H_{\mathrm{bulk}}
&= \sum^{L-1}_{j=1}
      \left[ -t~(c^{\dagger}_{j} c_{j+1}+ \mathrm{h.c.} )
      +\Delta~( c_{j} c_{j+1} + \mathrm{h.c.} ) \right] \nonumber\\
&~~~~~~- \sum^{L}_{j=1} \mu_{j} \left( c_{j}^{\dag} c_{j} - \frac{1}{2} \right) ,
      \label{eq:Ham1Bulk}
\end{align}
where $t$ is the hopping amplitude and $\Delta$ is the $p$-wave pairing gap, both of which can be assumed to be nonnegative without loss of generality. Here, $\mu_{j}$ is the on-site (chemical) potential, and we set those on the boundaries as $\mu_{1}=\mu_{L}={\sf a} \mu$ and those in the bulk as $\mu_{j}=\mu$ ($j=2,3,\cdots,L-1$), where ${\sf a} \ge 0$ is a constant. In the case of ${\sf a}=1$, $H_{\mathrm{bulk}}$ reduces to the Kitaev's $p$-wave superconductor model with open boundaries~\cite{Kitaev01}, in which Majorana edge zero modes occur provided that it is in the topological phase $\left| \mu/2t \right| \leq 1$. The boundary Hamiltonian is given by
\begin{equation}
H_{\mathrm{boundary}}
= {\sf b} \left[ -t~( e^{\ii \phi_{1}} c_{L}^{\dag} c_{1} + \mathrm{h.c.} )
+ \Delta~( e^{\ii \phi_{2}} c_{L} c_{1} + \mathrm{h.c.} ) \right],
\label{eq:Ham1Boundary}
\end{equation}
where $\phi_{1},\phi_{2} \in \left[ 0, 2\pi \right)$ are two independent phases that define the twisted boundaries, and ${\sf b} \ge 0$ is a constant. The TBC reduces to the open boundary condition (OBC) when ${\sf b}=0$. In the case of ${\sf a}={\sf b}=1$, the TBC boils down to the PBC for $\left( \phi_{1},\phi_{2} \right) = (0,0)$, or the APBC for $\left( \phi_{1},\phi_{2} \right) = (\pi,\pi)$.

Although the Hamiltonian $H$ does not conserve the total fermion number $F:= \sum_{j=1}^{L} c_{j}^{\dag} c_{j}$, the parity of the fermion number, i.e., the fermion number modulo $2$, is conserved since $H$ commutes with $P := (-1)^{F}$. Besides, time-reversal symmetry, i.e. invariance under complex conjugation, is respected if the chain has the periodic or anti-periodic boundaries.

It was pointed out in Ref.~\onlinecite{Greiter-14} that the fermionic parity of the ground state of $H$ with the PBC is 
odd in the topological phase.  
In order to see how the PBC is connected to the APBC through varying the twist parameters $(\phi_1,\phi_2)$, we diagonalize $H$ by the usual Fourier transform followed by a Bogoliubov transformation under the PBC and the APBC (see Appendix~\ref{APP_PAP}). The fermionic parity $P\mid_{(0,0)}$ or $P\mid_{(\pi,\pi)}$ in the ground state for the PBC or the APBC is summarized as follows:
\begin{enumerate}
\item Even $L$
\begin{center}
\begin{tabular}{ c | c  c } 
\ & $P\mid_{(0,0)}$ & $P\mid_{(\pi,\pi)}$  \\ \hline
$\left\lvert \mu/2t \right\rvert>1$ & $1$ & $1$ \\ 
$ \left\lvert \mu/2t \right\rvert<1$ & $-1$ & $1$ \\ 
\end{tabular}
\end{center}
\item Odd $L$
\begin{center}
\begin{tabular}{ c | c  c } 
\ & $P\mid_{(0,0)}$ & $P\mid_{(\pi,\pi)}$  \\ \hline
$\mu/2t >1$ & $-1$ & $-1$ \\ 
$-1 < \mu/2t <1$ & $-1$ & $1$ \\ 
$\mu/2t  <-1$ & $1$ & $1$ \\ 
\end{tabular}
\end{center}
\end{enumerate}

Thus, we always have
\begin{equation}
P\mid_{(0,0)}\times P\mid_{(\pi, \pi)}=
\begin{cases}
1 & \left\lvert \mu/2t \right\rvert>1,\\
-1 & \left\lvert \mu/2t \right\rvert<1,\\
\end{cases}
	\label{eq:SpecIntersection}
\end{equation}
regardless of whether $L$ is even or odd. The fact that $P\mid_{(0,0)}$ and $P\mid_{(\pi, \pi)}$ have opposite signs in the topological phase $\left\lvert \mu/2t \right\rvert<1$ indicates that, if we consider an evolution of parameters along a continuous path $\mathbf{\Phi} \left( s \right) =(\phi_1(s),\phi_2(s))\in \mathbb{R}^{2}$ with $s \in \left[ 0, 1 \right]$, such that $\mathbf{\Phi} \left( 0 \right) =\left( 0,0 \right)$ and $\mathbf{\Phi} \left( 1 \right) = \left( \pi,\pi \right)$, the ground state must be degenerate at some particular $s \in \left[ 0, 1 \right]$ since $P$ cannot be changed discontinuously without gap closing. As we will show below, the degeneracy of ground states actually indicates the existence of {\it Majorana zero modes}~\cite{Beenakker-13, Ortiz-14, Hegde-16}.

For each site $j$, we define the Majorana fermions by
\begin{equation}
a_{j} := c_{j} + c^{\dag}_{j},~~~b_{j} := (c_{j} - c^{\dag}_{j})/\ii.
\end{equation}
One can easily see that they satisfy the canonical anti-commutation relations for Majorana fermions. When written in terms of $a_{j}$ and $b_{j}$, the bulk Hamiltonian in Eq.~(\ref{eq:Ham1Bulk}) becomes
\begin{align}
H_{\mathrm{bulk}} &= \frac{\ii}{2} \sum^{L-1}_{j=1}
      \left[ (t+\Delta) b_{j} a_{j+1} - (t-\Delta) a_{j} b_{j+1} \right]
\nonumber \\
	&~~~~~- \frac{\ii}{2} \sum^L_{j=1} \mu_{j} a_{j} b_{j},
      \label{eq:Majo HamBulk}
\end{align}
and the Hamiltonian on the boundaries in Eq.~(\ref{eq:Ham1Boundary}) becomes
\begin{align}
&H_{\mathrm{boundary}}
\nonumber \\
&= \frac{\ii {\sf b}}{2} \left[ ( t \sin \phi_{1} \!-\! \Delta \sin \phi_{2} ) a_{1} a_{L}
\!+\! ( t \sin \phi_{1} \!+\! \Delta \sin \phi_{2} ) b_{1} b_{L} \right]
\nonumber \\
&~ - \frac{\ii {\sf b}}{2} \left[ ( t \cos \phi_{1} \!+\! \Delta \cos \phi_{2} ) a_{1} b_{L}
\!-\! ( t \cos \phi_{1} \!-\! \Delta \cos \phi_{2} ) b_{1} a_{L} \right].
      \label{eq:Majo HamBoundary}
\end{align}
We use in the following analysis the Majorana representation $a_{j}$ and $b_{j}$, instead of the ordinary fermionic representation $c_{j}$.

\section{Emergence of zero modes}
\label{sec:emergence of zero modes}

\begin{figure}[t]
\centering
\includegraphics[width=0.95\columnwidth]{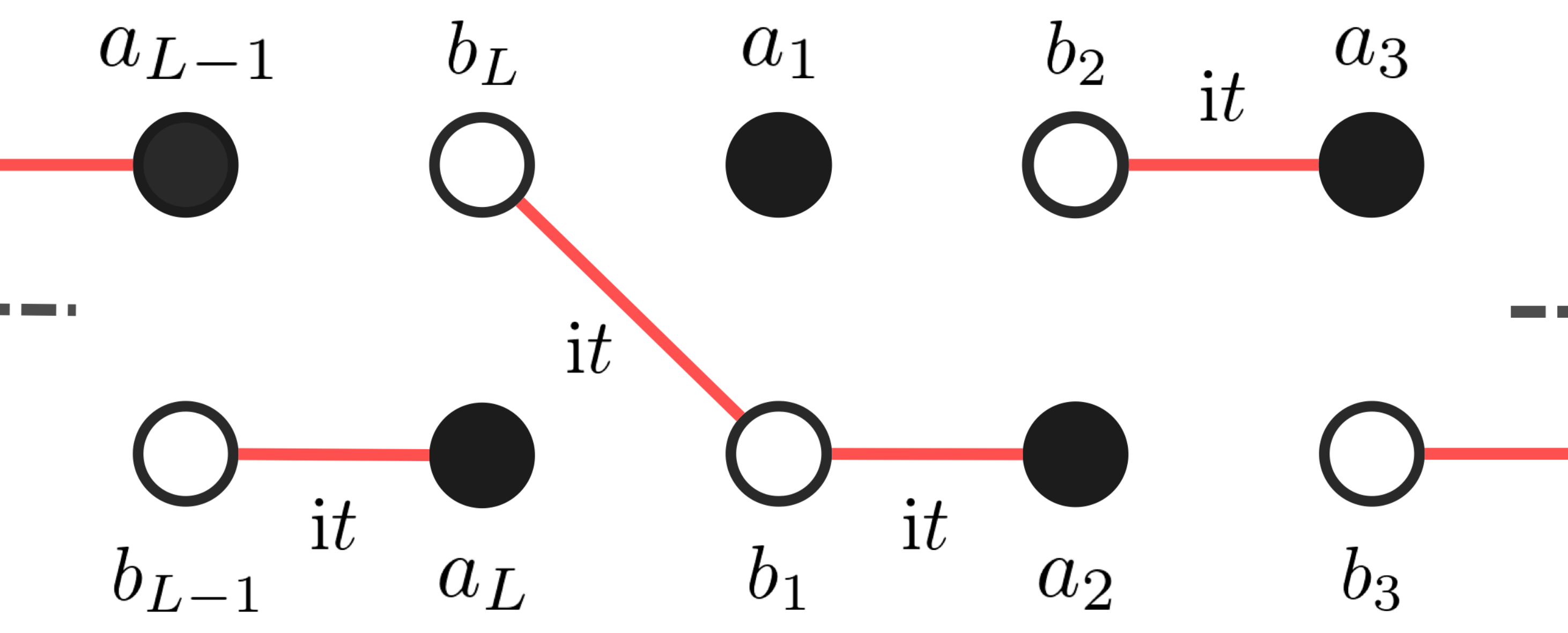}
\caption{(color online). 
The schematic representation of the chain with ${\sf a}={\sf b}=1$,~$\mu=0$,~$t=\Delta$, and $\phi_{1} = \phi_{2} = \pi/2$. The red lines represent the coupling between Majorana operators with the amplitude $\ii t$. The Majorana operator $a_1$ is isolated from the rest, and hence does not enter the Hamiltonian. The other Majorana operator that commutes with the Hamiltonian is $\left( b_{L} - a_{2} \right) /\sqrt{2}$.}
\label{fig:t=delta+mu0}
\end{figure}

In the previous section, it was demonstrated that the degeneracy of the ground states should happen between the PBC and the APBC, which implies the appearance of \textit{Majorana zero modes}. In this section, we show that the phase parameters $\left( \phi_{1},\phi_{2} \right)$ can be tuned so that the Majorana zero modes appear if and only if the system belongs to the topological phase. Using the expansion formula for Pfaffians, we explicitly calculate the parameter conditions for the existence of the zero modes for large chains. Note that Nava {\it et al}.~\cite{Giuliano-16} studied the Hamiltonian $H = H_{\rm bulk}+H_{\rm boundary}$ with $t=\Delta$, ${\sf a}=1$ in $H_{\rm bulk}$ and $\Delta=0$ in $H_{\rm boundary}$, and obtained the condition for the presence of Majorana zero modes using a different approach. Our method applies to the entire parameter region of the model and generalizes their results. 

In order to see how the Majorana zero modes emerge from the twisted boundary conditions, let us first consider the simplest case where ${\sf a}={\sf b}=1,~t=\Delta$, $\mu=0$, and $\phi_1=\phi_2=\pi/2$. In this case, the Hamiltonian is represented schematically in Fig.\ref{fig:t=delta+mu0}, and reduces to
\begin{align}
H = \ii t \sum_{j=2}^{L-1} b_{j} a_{j+1} + 
\ii t~b_{1} \left( a_{2} + b_{L} \right).
\end{align}
We see that the Majorana operator $a_{1}$ does not enter the Hamiltonian and thus corresponds to a Majorana zero mode. The other ``edge'' mode that commutes with the Hamiltonian is $\left( b_{L} - a_{2} \right) / \sqrt{2}$. We thus have a chain supporting zero modes even though there is \textit{no edge} in the system.

In general, the Hamiltonian given by Eqs. (\ref{eq:Majo HamBulk}) and (\ref{eq:Majo HamBoundary}) can be written in a quadratic form of Majorana fermions as
\begin{align}
H = \frac{\ii}{4} \sum_{i,j} d_{i} \left[ M_{L} \right]_{ij} d_{j}
	\label{eq:Majo QuadraticForm}
\end{align}
where  $d_{2i-1} := a_{i},~d_{2i} := b_{i}$ ($i=1,2, \cdots, L$). The ground state degeneracy  at some points on $\mathbf{\Phi} \left( s \right)$, mentioned in the preceding section,  implies the vanishing of $\det M_{L}$ at these points~\cite{Remark_InfiniteSystem}. The $2L\times2L$ real skew symmetric matrix $M_{L}$ can be expressed as
\begin{equation} 
M_{L}=\begin{pmatrix}
	{\sf a}\hat{m} & \hat{t}_{0} & \ & \ & {\sf b}\hat{t}_{1} \\
	-{\hat{t}_{0}}^{~T} & \hat{m} & \ & \ & \ \\
	\ & \ddots & \ddots & \ddots & \ \\
	\ & \ & \ & \hat{m} & {\hat{t}_{0}} \\
	-{\sf b} {\hat{t}_{1}}^{~T} & \ & \ & -{\hat{t}_{0}}^{~T} & {\sf a} \hat{m} \\
\end{pmatrix}, 
\end{equation}
where the empty entries are zero and 
\begin{eqnarray}
\hat{m} \! &:=& \! 
\begin{pmatrix}
	0 & -\mu \\
	\mu & 0 \\
\end{pmatrix}, \\
\hat{t}_{0} \! &:=& \! 
\begin{pmatrix}
	0 & -(t-\Delta) \\
	t+\Delta & 0 \\
\end{pmatrix},\\
\hat{t}_{1} \!  &:=&\!  
\begin{pmatrix}
	t\sin\phi_{1} \!-\! \Delta\sin\phi_{2} & -t\cos\phi_{1} \!+\! \Delta\cos\phi_{2} \\
	t\cos\phi_{1} \!+\! \Delta\cos\phi_{2} & t\sin\phi_{1} \!+\! \Delta\sin\phi_{2} \\
\end{pmatrix}.
\end{eqnarray}

\onecolumngrid
\medskip
Using the expansion formula (see Sec. 2.8 of Ref.~\onlinecite{Laplace expansion}), the Pfaffian of $M_L$ can be obtained as
\begin{align}
\begin{split}
\mathrm{Pf}~M_{L}
=&\mathrm{Pf}~\tilde{M}_{L}^{\mathrm{(open)}} + {\sf b}^{2} \left( t^2 -\Delta^2 \right) \mathrm{Pf}~\tilde{M}_{L-2}^{\mathrm{(open)}} \\
&- {\sf b} \left( t \cos \phi_{1} + \Delta \cos \phi_{2} \right) \left( t + \Delta \right)^{L-1}
+ {\sf b} \left( t \cos \phi_{1} - \Delta \cos \phi_{2} \right) \left(- ( t - \Delta ) \right)^{L-1},
\end{split}
	\label{eq:PfExplicit}
\end{align}
where $\tilde{M}^{\left( \mathrm{open} \right)}_{L} := M_{L}|_{{\sf b}=0}$ is the matrix representation for the chain with open boundaries (but with arbitrary ${\sf a}$).
Since the determinant of $M_L$ is related to $\mathrm{Pf}~M_{L}$ via $\det M_{L}= \left( \mathrm{Pf}~M_{L} \right)^{2}$,
we impose $\mathrm{Pf}~M_{L}=0$, yielding
\begin{equation}
\begin{split}
{\sf b} \left( t \cos \phi_{1} + \Delta \cos \phi_{2} \right) 
- {\sf b} \left( t \cos \phi_{1} - \Delta \cos \phi_{2} \right) \left( - \frac{t-\Delta}{t+\Delta} \right)^{L-1}
=\left( t+\Delta \right) \frac{\mathrm{Pf}~\tilde{M}_{L}^{\mathrm{(open)}}}{\left( t+\Delta \right)^{L}} 
+ {\sf b}^{2} \left( t-\Delta \right) \frac{\mathrm{Pf}~\tilde{M}_{L-2}^{\mathrm{(open)}}}{\left( t+\Delta \right)^{L-2}}.
\end{split}
	\label{eq:rhsconverge}
\end{equation}

\begin{figure*}[t]
\centering
\includegraphics[width=0.95\columnwidth]{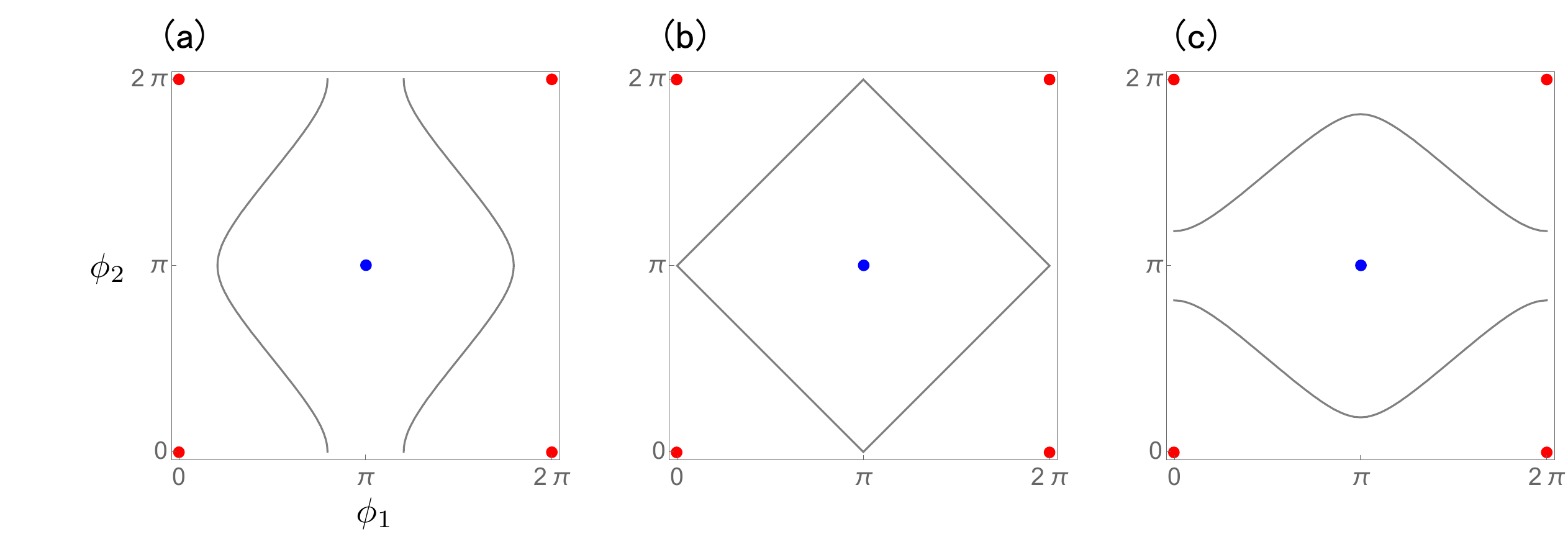}
\caption{(color online). Examples of the zeros of $\det M$ for large chains. The solution of $\det M=0$, i.e., the solution of $t \cos \phi_{1} + \Delta \cos \phi_{2} = 0$, is represented on a $\phi_{1}$-$\phi_{2}$ plane ($\phi_{1}, \phi_{2} \in \left[ 0, 2\pi \right)$, grey curves), which separates $(0, 0)$ (red points) and $(\pi,\pi)$ (blue points).  (a) $\Delta = 0.8t$, (b) $\Delta = t$, (c) $\Delta = 1.2t$.}
	\label{fig:PfZero}
\end{figure*}

\twocolumngrid

When $L$ is large, the phase parameters $\left( \phi_{1}, \phi_{2} \right)$ that exactly satisfy $\mathrm{Pf}~M_{L} = 0$ exist only when the right hand side of Eq.~(\ref{eq:rhsconverge}) converges. Note that $|(t-\Delta)/(t+\Delta)| \le 1$, because both $t$ and $\Delta$ are nonnegative. From the recurrence relation of $\mathrm{Pf}~M^{\mathrm{ \left( open \right)}}_{L}$,
\begin{align}
&\mathrm{Pf}~\tilde{M}_{L}^{\left( \mathrm{open} \right)} \nonumber \\
&~~~= -\mu \mathrm{Pf}~\tilde{M}_{L-1}^{\left( \mathrm{open} \right)} - \left( t^{2} - \Delta^{2} \right) \mathrm{Pf}~\tilde{M}_{L-2}^{\left( \mathrm{open} \right)},
	\label{eq:Recurrence OBC}
\end{align}
the asymptotic form of $\mathrm{Pf}~\tilde{M}_{L}^{\left( \mathrm{open} \right)}$ for large $L$ is obtained as
\begin{align}
\mathrm{Pf}~\tilde{M}_{L}^{\left( \mathrm{open} \right)}
\sim \Lambda^{L},
	\label{eq:Asymp OBC}
\end{align}
where 
\begin{equation}
\Lambda = -\frac{\mu}{2} \left( 1 + \sqrt{1- \frac{4 \left( t^{2} - \Delta^{2} \right)}{\mu^{2}}} \right)
\end{equation}
is the solution of the characteristic equation of the recurrence relation Eq.~(\ref{eq:Recurrence OBC}) with the largest absolute value. 
Then, the condition that the right hand side of Eq.~(\ref{eq:rhsconverge}) converges for large $L$ is given by
\begin{equation}
\left| \frac{\Lambda}{t+\Delta} \right| < 1,
	\label{ConvergeCondition}
\end{equation}
which just reduces to the condition for the topological phase:
\begin{align}
\left\lvert\frac{\mu}{2t}\right\rvert<1.
\label{eq:Topophase}
\end{align}
In this case, Eq.~(\ref{eq:rhsconverge}) becomes
\begin{align}
{\sf b} \left( t+\Delta \right)^{L-1}\left[ \left( t\cos\phi_1+\Delta\cos\phi_2 \right) - \varepsilon \left( L \right) \right] = 0,
	\label{eq:PfZero}
\end{align}
where
\begin{align}
\varepsilon(L)= O \left(\left(\frac{t-\Delta}{t+\Delta}\right)^{L-1}\right)+ O \left(\left(\frac{\Lambda}{t+\Delta}\right)^{L-1}\right)
	\label{eq:eps_L}
\end{align}
is a small term for sufficiently large $L$. Thus, Eq.~(\ref{eq:PfZero}) gives an $L$-dependent solution $\left( \phi_1 \left( L \right),\phi_2 \left( L \right) \right)$ of $\mathrm{Pf}~M_{L}=0$.

In the case of $\phi_{1} = \phi_{2}$, Eq.~(\ref{eq:PfZero}) reduces to $\phi_{1} = \phi_{2} \simeq \pi/2$, $3\pi/2$~[\onlinecite{Remark_Duality_Defect}]. 
Since $\left( \phi_{1}, \phi_{2} \right)=(0,0)$ and $(\pi,\pi)$ are separated by the curve described by Eq.~(\ref{eq:PfZero}) on a $\phi_{1}$-$\phi_{2}$ plane (see Fig.\ref{fig:PfZero}), an arbitrary path ${\bm \Phi}(s)$ from $(0,0)$ to $(\pi,\pi)$ must intersect the curve when the system belongs to the topological phase. 
Note that Eq.~(\ref{eq:PfZero}) applies to arbitrary boundary conditions, while Eq.~(\ref{eq:SpecIntersection}) is valid only for the PBC and the APBC. Therefore, the intersection between ${\bm \Phi}(s)$ and the curve Eq.~(\ref{eq:PfZero}) can be viewed as a general condition which characterizes the topological phase.

\section{Zero modes for $\phi_{1} = \phi_{2} = \pi/2$}
\label{sec:exact zero modes}

So far we have confirmed the presence of Majorana zero modes in large chains belonging to the topological phase. However, the spatial profile of the zero modes is not obtained and it is unclear whether the zero modes are localized. In this section, we discuss the explicit forms of the zero modes and their properties for finite chains with $\phi_{1} = \phi_{2} = \pi/2$. We first focus on a simple discussion by the Chebyshev polynomials and then determine explicit forms of the zero modes. Moreover, we present the full spectrum of the chain with $\mu=0$, and finally, we demonstrate that the topological order survives even in the presence of disorder.

\subsection{Conditions for the presence of exact zero modes by the Chebyshev polynomials}
\label{sec:Chebyshev}

\begin{figure}[t]
		\includegraphics[width=0.95\columnwidth]{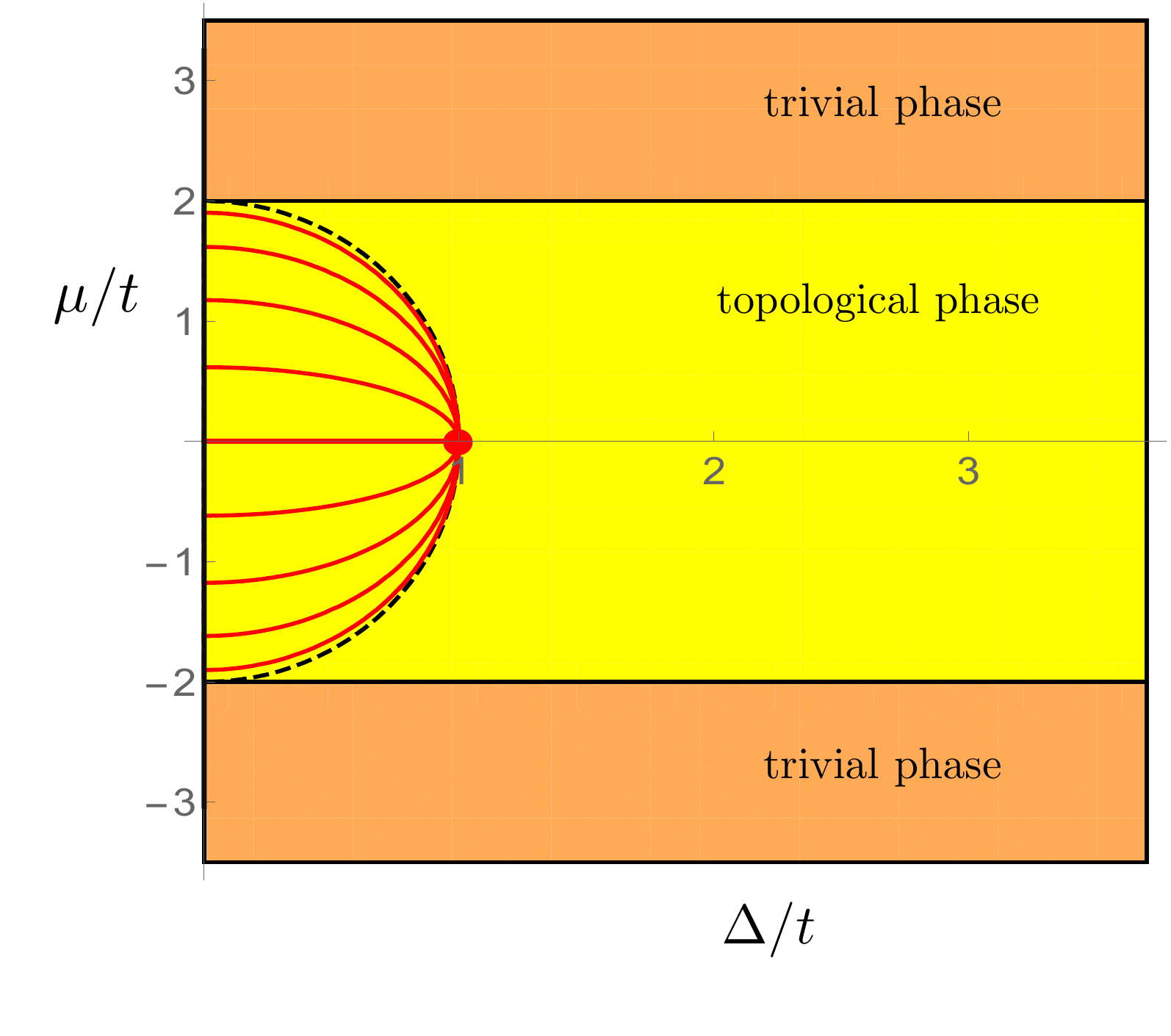} 
      \caption{(color online). Phase diagram of the Majorana chain with open boundaries. The solution curves of Eq.~(\ref{zero modes, OBC}) with $L=9$ are represented by red curves, which exist only in the region $(\Delta/t)^2+(\mu/2t)^2\le1$. Outside the region, exact zero modes never appear even if the chain belongs to the topological phase. }
      \label{fig:phase diagram OBC}
\end{figure}

First of all, we consider the Majorana chain with open boundaries. We here set ${\sf a}=1$ in particular. In this case, the following recurrence relations and initial conditions hold:
\begin{align}
\mathrm{Pf}~M_{L}^{\left( \mathrm{open} \right)} = -\mu \mathrm{Pf}~M_{L-1}^{\left( \mathrm{open} \right)} - \tau^{2} \mathrm{Pf}~M_{L-2}^{\left( \mathrm{open} \right)}
\label{eq:Recurrence OBC2} \\
\mathrm{Pf}~M_{1} ^{\left( \mathrm{open} \right)} = -\mu,~
\mathrm{Pf}~M_{2} ^{\left( \mathrm{open} \right)} = \mu^{2} - \tau^{2},
\end{align}
where we define $\tau := \sqrt{t^{2} - \Delta^{2}}$, which can be either real or imaginary depending on the sign of $t^2-\Delta^2$. Note that $\mathrm{Pf}~M_{1}^{\left( \mathrm{open} \right)},~\mathrm{Pf}~M_{2}^{\left( \mathrm{open} \right)}$ are defined so that they are compatible with the recurrence relation Eq.~(\ref{eq:Recurrence OBC2}) and $\mathrm{Pf}~M_{3}^{\left( \mathrm{open} \right)} ,~\mathrm{Pf}~M_{4}^{\left( \mathrm{open} \right)} $. 
Because these recurrence relations and initial conditions are the same as those of the Chebyshev polynomials of the second kind $U_{L} \left( z \right)$~\cite{Definition_Chebyshev}, we could express $\mathrm{Pf}~M_{L}^{\left( \mathrm{open} \right)}$ as
\begin{equation}
\mathrm{Pf}~M_{L}^{\left( \mathrm{open} \right)} = \tau^{L} \cdot U_{L} \left( - \frac{\mu}{2 \tau} \right).
\end{equation}
The necessary and sufficient condition for the presence of exact zero modes in finite chains is $\mathrm{Pf}~M_{L}^{\left( \mathrm{open} \right)} = 0$. Since $U_{L} \left( z \right)$ has zeros only in the interval $z \in \left[ -1,1 \right]$, exact zero modes for finite $L$ appear for $\mu^{2} < 4 \left( t^2-\Delta^2 \right)$ and on the curves
\begin{equation}
\frac{\mu}{t} = 2 \sqrt{1 - \left(\frac{\Delta}{t}\right)^{2}} \cos \frac{k \pi}{L+1}~~\left( k = 1,2, \cdots, L \right)
	\label{zero modes, OBC}
\end{equation}
in the $\Delta/t$-$\mu/t$ phase diagram (see Fig.~\ref{fig:phase diagram OBC}), reproducing the results in Ref.~\onlinecite{Kao-14}. For the remaining regions $\mu^{2} > 4 \left( t^2-\Delta^2 \right)$, 
exact zero modes are absent even when the chain belongs to the topological phase $\left| \mu/2t \right| < 1$~\cite{Kao-14}. The following discussions on the Majorana chains with twisted boundaries become simple because $\mathrm{Pf}~M_{L}^{\left( \mathrm{open} \right)}$ can be expressed as the Chebyshev polynomials~\cite{Wakatsuki-14}.

For a chain with twisted boundaries, especially with $\phi_{1} = \phi_{2} = \pi/2$, we find the following relation between the Pfaffian under the TBC and that under the OBC,
\begin{eqnarray}
\mathrm{Pf}~M_{L}
&=& {\sf a}^{2} \mathrm{Pf}~M_{L}^{\left( \mathrm{open} \right)}
-[ 2 {\sf a} \left( 1 - {\sf a} \right) + {\sf b}^{2}] \tau^{2} \mathrm{Pf}~M_{L-2}^{\left( \mathrm{open} \right)}\nonumber\\
&&+ \left( {\sf a} - 1 \right)^{2} \tau^{4} \mathrm{Pf}~M_{L-4}^{\left( \mathrm{open} \right)},
\end{eqnarray}
which leads to the condition that exact zero modes appear in finite chains ($z := -\mu/2\tau$)
\begin{equation}
U_{L} \left( z \right)
- \frac{2 {\sf a} \left( 1 - {\sf a} \right) + {\sf b}^{2}}{{\sf a}^{2}} U_{L-2} \left( z \right)
+ \left( \frac{{\sf a} - 1}{{\sf a}} \right)^{2} U_{L-4} \left( z \right) = 0.
\label{eq:pre_zeromode}
\end{equation}
Letting $z =: \cos \theta$ ($\theta \in \mathbb{C}$), a simple calculation gives
\begin{equation}\label{zeromode_condition}
\left( \cos^{2} \theta - \frac{2 {\sf a} + {\sf b}^{2}}{{\sf a}^{2}} \right) \sin \left[ \left( L-1 \right) \theta \right] + \frac{1-2 {\sf a}}{4 {\sf a}} \sin \left[ \left( L -3 \right) \theta \right] = 0.
\end{equation}

It seems difficult to solve this equation in general, but some simplification occurs in the following cases: 
\begin{enumerate}
\item ${\sf a}={\sf b}=1$

In this case, Eq.~(\ref{zeromode_condition}) reduces to
\begin{equation}
2 \cos L \theta = 0,
\end{equation}
with solutions
\begin{equation}
z = \cos \frac{\left( 2k-1 \right) \pi}{2L}~~\left( k = 1,2, \cdots, L \right).
\end{equation}

\item ${\sf a} = 1/2$

In this case, the condition Eq.~(\ref{zeromode_condition}) reduces to
\begin{equation}
\sin [ \left( L - 1 \right) \theta ] \left[ \cos^{2} \theta - \left( 1 +  {\sf b}^{2} \right)  \right] = 0,
\end{equation}
with solutions (for $ k = 1,2, \cdots, L-2$)
\begin{equation}
z = \pm \sqrt{1+ {\sf b}^{2}},~\cos \frac{k \pi}{L-1},
\end{equation}
indicating that exact zero modes appear for \textit{all} $L$ for the chain with ${\sf a}=1/2$ and $\mu = \pm 2\sqrt{t^2-\Delta^2}\sqrt{1+{\sf b}^{2}}$. We can also prove the contrary, that is, the necessary condition that exact zero modes appear for all $L$ is ${\sf a}=1/2,~\mu = \pm 2\sqrt{t^2-\Delta^2}\sqrt{1+{\sf b}^{2}}$, except for the trivial case $t=\Delta,~\mu=0$. 
To prove this, we first note that $\mathrm{Pf}~M_{L}$ 
obeys the recurrence relation for $L \geq 5$, which is the same as Eq.~(\ref{eq:Recurrence OBC2}). 
We set 
\begin{align}
&\mathrm{Pf}~M_{1} = \left( 1-2{\sf a} \right) \mu,~ \nonumber \\
&\mathrm{Pf}~M_{2} = {\sf a}^{2} \mu^{2} - \left( 1+{\sf b}^{2} \right) \tau^{2},
\end{align}
so that they are compatible with the recurrence relation 
and $\mathrm{Pf}~M_{3},~\mathrm{Pf}~M_{4}$. 
Then, if the exact zero modes exist for all $L$ with the specific parameters, both  $\mathrm{Pf}~M_{1}$ and $\mathrm{Pf}~M_{2}$ are required to be zero by the Euclidean algorithm, which completes the proof.

\item $\mu=0$

If $L$ is odd, all of $U_{L} \left( 0 \right), U_{L-2} \left( 0 \right)$, and $U_{L-4} \left( 0 \right)$ are zero and the exact zero mode condition is always satisfied, which leads to \textit{ever-presence} of the exact zero modes. On the other hand, if $L$ is even, the identities $U_{L} \left( 0 \right) = -U_{L-2} \left( 0 \right) = U_{L-4} \left( 0 \right)$ lead to ${\sf b}^{2} + 1 = 0$, which implies \textit{never-presence} of the exact zero modes. Actually, the full spectrum of the Hamiltonian can be determined analytically for $\mu=0$, as will be presented in \ref{sec:full spectrum mu0}.

\item $\mu = \pm 2 \tau \left( \neq 0 \right)$

This case is of particular importance in the discussion of solvable interacting Majorana chains~\cite{Katsura_Int_Majorana}, where $\mu = \pm 2 \tau$ is just the frustration-free condition in the noninteracting limit. In this case, the identities $U_{L} \left( 1 \right) = L+1$ and $U_{L} \left( -1 \right) = \left( -1 \right)^{L} \left( L+1 \right)$ give
\begin{eqnarray}
0
&=& \left( 2 {\sf a} + {\sf b} - 1 \right) \left( 2 {\sf a} - {\sf b} - 1 \right) L \nonumber \\
&~&~~~~~~~~~~~~~ - \left( 2 {\sf a} - 1 \right) \left( 2 {\sf a} - 3 \right) + {\sf b}^{2},
\end{eqnarray}
which leads to the necessary condition ${\sf a}=1/2$ and ${\sf b}=0$ for the chain to have exact zero modes for arbitrary $L$. In other words, exact zero modes do not appear when $\mu = \pm 2 \tau$ 
and at the same time the chain has boundary terms.
\end{enumerate}

\subsection{Explicit forms of zero modes}
\label{sec:expform}

In this subsection, we show explicit forms of the zero modes. A zero mode $\bm{\Psi_{0}}$ is expressed as $\left( \bm{\Psi_{0}} \right)_{2i-1} = A_{i},~\left( \bm{\Psi_{0}} \right)_{2i} = B_{i}$ ($i=1,2,\cdots,L$) and satisfies the following relations for the bulk ($i=1,2, \cdots , L$)
\begin{eqnarray}
\left( t - \Delta \right) A_{i-1} + \mu A_{i} + \left( t + \Delta \right) A_{i+1} &=& 0 \nonumber \\
\left( t + \Delta \right) B_{i-1} + \mu B_{i} + \left( t - \Delta \right) B_{i+1} &=& 0,
	\label{zero modes bulk}
\end{eqnarray}
where we have introduced four virtual variables $A_{0}, B_{0}, A_{L+1}$, and $B_{L+1}$ that can be incorporated into the following boundary conditions
\begin{eqnarray}
\left( t+\Delta \right) B_{0} + \tilde{\sf a} \mu B_{1} + {\sf b} \left( t - \Delta \right) A_{L} &=& 0 \nonumber \\
- \left( t-\Delta \right) A_{0} - \tilde{\sf a} \mu A_{1} + {\sf b} \left( t+\Delta \right) B_{L} &=& 0 \nonumber \\
- {\sf b} \left( t - \Delta \right) A_{1} + \tilde{\sf a} \mu B_{L} + \left( t-\Delta \right) B_{L+1} &=& 0 \nonumber \\
-{\sf b} \left( t + \Delta \right) B_{1} - \tilde{\sf a} \mu A_{L} - \left( t+\Delta \right) A_{L+1} &=& 0,
\label{bc}
\end{eqnarray}
where $\tilde{\sf a} := 1 - {\sf a}$.

The bulk conditions form second-order linear recurrence equations whose general solutions can be written as $A_{i} = A_{+} \lambda_{+}^{i} + A_{-} \lambda_{-}^{i},~B_{i} = B_{+} \lambda_{+}^{L-i+1} + B_{-} \lambda_{-}^{L-i+1}$, with
\begin{align}
\lambda_{\pm} := \frac{- \mu \pm \sqrt{\mu^{2} - 4 \left( t^{2} - \Delta^{2} \right)}}{2 \left( t + \Delta \right)}.
\end{align}
Note that the absolute value of $\lambda_{+}$ and $\lambda_{-}$ must be less than $1$ in order for the zero modes to have finite normalization for large $L$, which implies the topological condition Eq.~(\ref{eq:Topophase}). The coefficients $A_\pm$ and $B_\pm$ are determined by the boundary conditions Eq.~(\ref{bc})
\begin{equation}
X \left( A_{+} ~ A_{-} ~ B_{+} ~ B_{-} \right)^{T} = 0,
	\label{zero mode matrix}
\end{equation}
where $X$ is a $4 \times 4$ matrix defined by the parameters (see Appendix \ref{sec:zero mode matrix} for the specific form of $X$). Therefore, the necessary and sufficient condition for the existence of the zero modes is $\mathrm{det}~X = 0$. We here remark that this condition can be applied to infinite systems as well as finite systems. In fact, for infinite chains, it is clear that $\det X$ will always become zero when the systems belong to the topological phase Eq.~(\ref{eq:Topophase}). Therefore, we again confirm the presence of the zero modes for large chains in the topological phase. By their explicit forms, it is evident that the zero modes are localized if they exist.

The condition obtained can be simplified in some specific cases. First, for a chain with open boundaries (${\sf a}=1,~{\sf b}=0$), $M$ becomes block diagonal and
$\det X = 0$ reduces to $\left( \lambda_{+} / \lambda_{-} \right)^{L+1} = 1$, which recovers Eq.~(\ref{zero modes, OBC}) obtained using the Chebyshev polynomials. Next, for a chain with $t=\Delta$, $\det X = 0$ is \textit{always} satisfied, which confirms the existence of exact zero modes for the chain in this case regardless of the other conditions.

Finally, in the case of the chain with $\mu=0$, we have
\begin{equation}
\det X \propto \left( 1 + \left( -1 \right)^{L} \right)^{2} \left( \frac{t-\Delta}{t+\Delta} \right)^{L}.
\end{equation}
The exact zero modes thus appear when the chain length $L$ is odd regardless of other conditions. For even $L$, exact zero modes never appear unless $t=\Delta$. However, as shown above, there always exist the zero modes (though not an \textit{exact} one) in the thermodynamic limit. All of these observations are consistent with the previous results presented in \ref{sec:Chebyshev}. We will discuss the case of $\mu=0$ in more detail in the next subsection.

\subsection{Full spectrum for $\mu=0$}
\label{sec:full spectrum mu0}
We have discussed the conditions for the existence of zero modes in the case of $\phi_{1} = \phi_{2} = \pi/2$. Besides the zero modes, we are also interested in determining the \textit{full spectrum} of the chain. In particular, the chain with $\mu = 0$ can be \textit{solved exactly} as follows.
We set $J := t+\Delta,~f := t-\Delta$ and assume that $J \neq 0$ and $f \neq 0$. 
Then the Hamiltonian with $\mu=0$ reads
\begin{equation}
H = \frac{\ii}{2} \left[ \sum_{j=1}^{L-1} \left( J b_{j} a_{j+1} - f a_{j} b_{j+1} \right) + {\sf b} \left( f a_{1} a_{L} + J b_{1} b_{L} \right) \right].
\end{equation}
When the energy eigenvalues of $H$ are expressed as $E = \varepsilon /4$, the conditions for the bulk are ($i=1,2, \cdots, L$)
\begin{eqnarray}
\ii fA_{i-1} + \ii JA_{i+1} &=& \varepsilon B_{i} \nonumber \\
- \ii JB_{i-1} - \ii fB_{i+1} &=& \varepsilon A_{i},
	\label{mu0 bulk condition}
\end{eqnarray}
and the conditions for the boundary are
\begin{eqnarray}
JB_{0} + {\sf b} f A_{L} &=& 0 \nonumber \\
fA_{0} - {\sf b} J B_{L} &=& 0 \nonumber \\
{\sf b} A_{1} - B_{L+1} &=& 0 \nonumber \\
{\sf b} B_{1} + A_{L+1} &=& 0,
	\label{mu0 boundary conditions}
\end{eqnarray}
where the virtual variables $A_{0}, B_{0}, A_{L+1}, B_{L+1}$ are defined by the bulk conditions Eq.~(\ref{mu0 bulk condition}). Since the spectrum is chiral, we concentrate on the nonnegative eigenvalues $E\geq 0$ in what follows.

\begin{figure}[t]
      \begin{minipage}[t]{0.95\hsize}
		\centering
		\includegraphics[width=0.95\columnwidth]{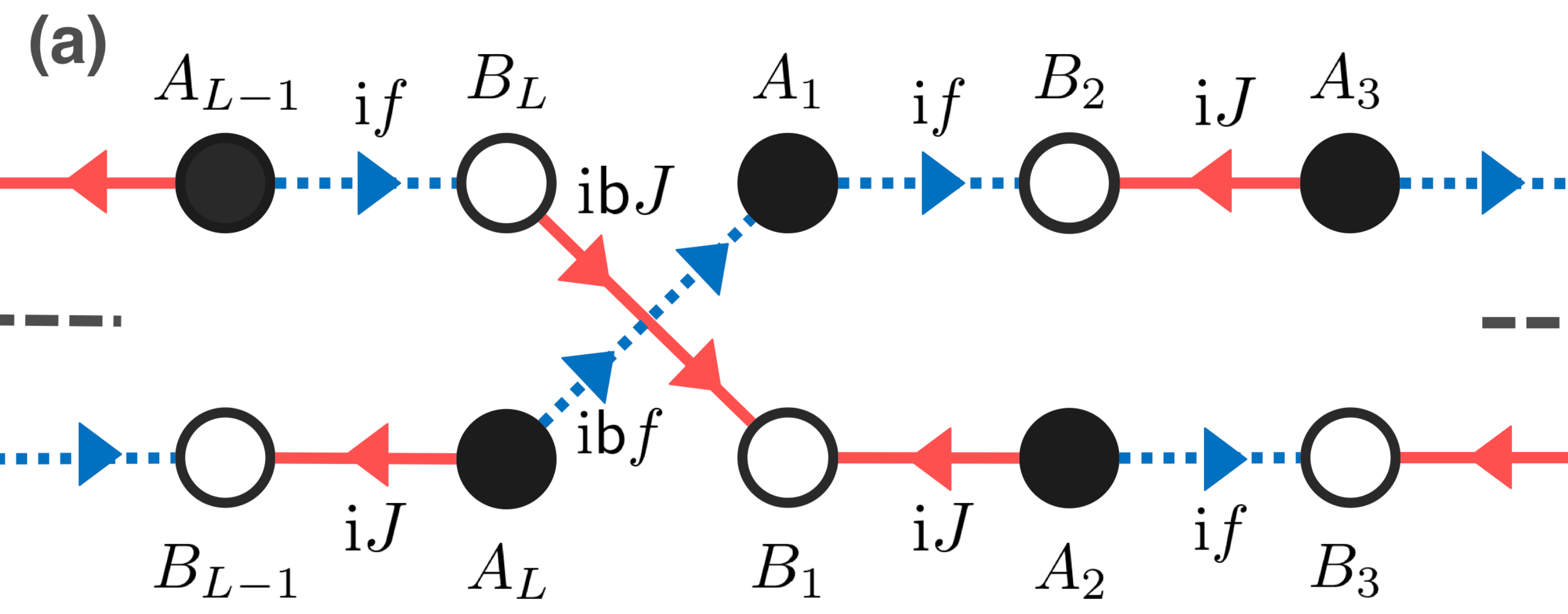}
      \end{minipage} \\ 
      \begin{minipage}[t]{0.95\hsize}
		\centering
		\includegraphics[width=0.95\columnwidth]{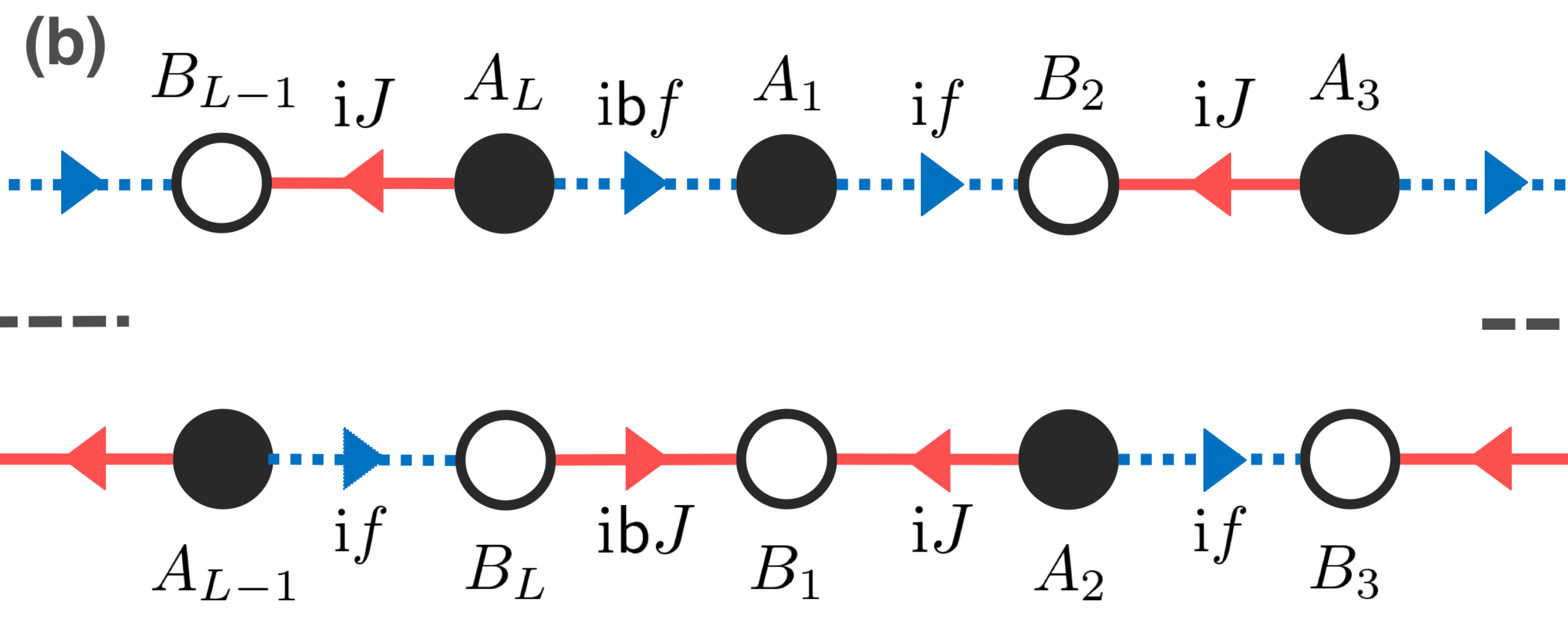}
      \end{minipage}
      \caption{(color online). The schematic representation of the chain with $\mu=0$. (a) Even $L$: one M\"obius ring with two defects. (b) Odd $L$: two decoupled chains, each of which has one defect.}
      \label{fig:mu0}
\end{figure}

The structure of the Hamiltonian depends on the parity of $L$. When $L$ is even, $H$ forms a nearest-neighboring closed chain (just like a \textit{M\"obius ring}) of length $2L$ with two defects (Fig.~\ref{fig:mu0}a)~\cite{PLA}. On the other hand, if $L$ is odd, the Hamiltonian is separated into two \textit{decoupled} chains with nearest-neighbor hopping, each of which is closed and has one defect bond (Fig.~\ref{fig:mu0}b). 

\begin{figure}[t]
	\includegraphics[width=0.85\columnwidth]{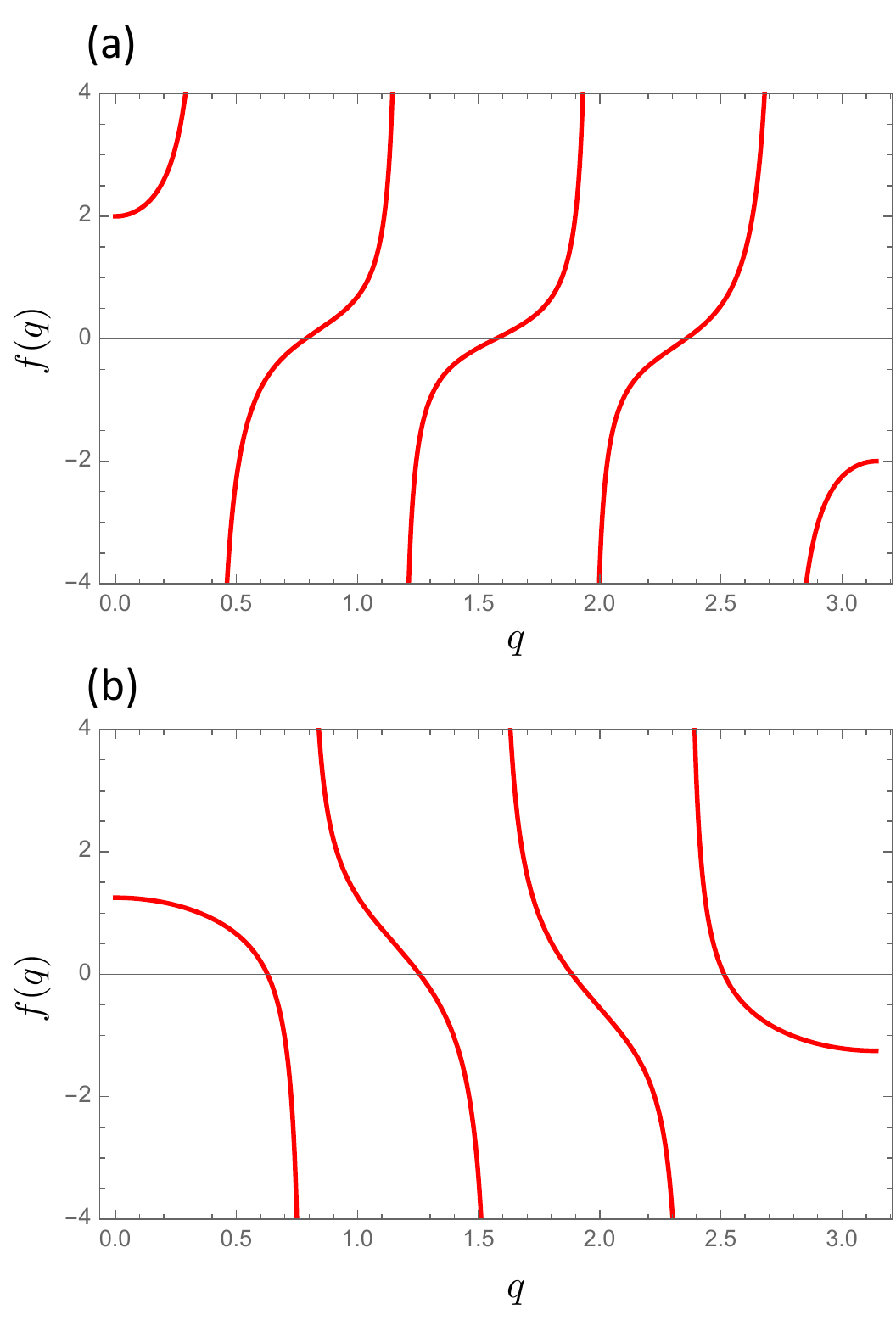} 
      \caption{The graphic representation of the functions which determine the quantized wavenumbers corresponding to the eigenvalues. The allowed wavenumbers are obtained by the intersection between the above graphs and the constant calculated by the parameters. (a) Even $L$: $f \left( q \right) = \sin qM / \left[ \sin q \left( M+1 \right) - {\sf b}^{2} \sin q \left( M-1 \right) \right]$ for $L=8$ and ${\sf b}=1$. (b) Odd $L$: $f \left( q \right) = \sin qN / \sin q \left( N-1 \right)$ for $L=9$.}
      \label{fig:graph}
\end{figure}

For even $L$, the solutions on the foregoing $2L$-chain are obtained by conducting a plane-wave expansion independently in two parts of the M\"obius ring and combining them at the two defects. The exact eigenvalues are given by (see Appendix \ref{sec:mu0 detail})
\begin{equation}
\varepsilon = \pm \sqrt{J^{2} + f^{2} + 2Jf \cos q},
	\label{dispersion relation}
\end{equation}
where the wave numbers $q$ are determined by the following quantization condition ($M:=L/2$, Fig.~\ref{fig:graph}a):
\begin{eqnarray}
\frac{\sin qM}{\sin q \left( M+1 \right) - {\sf b}^{2} \sin q \left( M-1 \right) } \nonumber \\
~~~~~= \frac{Jf}{{\sf b}^{2}J^{2} - f^{2}}~~\mathrm{or}~~\frac{Jf}{{\sf b}^{2}f^{2} - J^{2}}.
	\label{quantization condition for even}
\end{eqnarray}
For $\Delta \neq 0$, the existence of zero modes is only possible for the wavenumber $q$ with nonvanishing imaginary part since the zero energy is out of the energy band. 
It follows from the reality of $\varepsilon$ that $q$ takes the form 
$q = m\pi + \ii q_{*}$, where $m \in \mathbb{Z}$ and $q_{*} \in \mathbb{R}$.
Thus, the dispersion relation reads 
\begin{equation}
\varepsilon = \pm \sqrt{J^{2} + f^{2}+2 \left( - 1 \right)^{m} Jf \cosh q_{*}},
	\label{dispersion relation2}
\end{equation}
and the quantization condition given by Eq.~(\ref{quantization condition for even}) in the case of \emph{large} $L$ will be
\begin{eqnarray}
\frac{ (-1)^{m} }{e^{q_{*}} - {\sf b}^2 e^{-q_{*}}} \simeq \frac{Jf}{{\sf b}^{2}J^{2} -f^{2}}~~\mathrm{or}~~\frac{Jf}{{\sf b}^{2} f^{2}-J^{2}}.
\end{eqnarray}
After a straightforward calculation, we have an eigenvalue which is exponentially small for large even $L$:
\begin{equation}
\varepsilon
\simeq \sqrt{\frac{1+{\sf b}^{2}}{J^{2} + {\sf b}^{2}f^{2}}} \left( J^{2} - f^{2} \right) \left( \frac{f}{J} \right)^{L/2}.
\end{equation}

The analysis for odd $L$ can be performed in a similar way. The dispersion relation is the same as Eq.~(\ref{dispersion relation}), and the wave numbers $q$ are determined by the following quantization condition ($N:=(L+1) / 2$, Fig.~\ref{fig:graph}b):
\begin{equation}
e^{\ii q} = - \frac{f}{J}~~\mathrm{or}~~\frac{\sin qN}{\sin q \left( N-1 \right)} = \left\{ \frac{{\sf b}^{2} J}{f}~~\mathrm{or}~~\frac{{\sf b}^{2} f}{J} \right\}.
	\label{quantization condition for odd}
\end{equation}
The wavenumber determined by the first equality corresponds to the exact zero mode. Here, due to the fact that the spectrum is chiral and the number of eigenvalues is odd, there always exist exact zero modes for odd $L$. In fact, the Majorana zero operators are
\begin{eqnarray}
\Psi_{1} &=& a_{1} + \sum_{m=1}^{N-1}  \left( a_{2m+1} + {\sf b}\,  b_{2N-2m} \right) \left( - \frac{f}{J} \right)^{m}, \label{mu0 zero mode1} \\
\Psi_{2} &=& \left( b_{2N-1} - {\sf b}\, a_{2} \right)
+ \sum_{m=1}^{N-2}~( b_{2N-2m-1} \nonumber \\ 
&-& {\sf b}\, a_{2m+2} ) \left( - \frac{f}{J} \right)^{m} + \left( - \frac{f}{J} \right)^{N-1} b_{1}. \label{mu0 zero mode2}
\end{eqnarray}
Since we assumed from the outset that $|f/J|<1$, the coefficients fall off exponentially in system size and hence each operator has finite normalization even in the $L \to \infty$ limit, i.e. $(\Psi_i)^2 ={\rm const.}<\infty$ ($i=1,2$). 

When $L$ is large, the quantization condition Eq.~(\ref{quantization condition for odd}) except for that of the exact zero mode becomes 
\begin{equation}
\left( -1 \right)^{m} e^{q_{*}}\simeq \frac{{\sf b}^{2} J}{f}~~\mathrm{or}~~\frac{{\sf b}^{2} f}{J},
	\label{quantization condition for odd_largeL}
\end{equation}
where we have substituted $q = m\pi+\ii q_{*}$. 
Equation (\ref{quantization condition for odd_largeL}) does not have a solution because  it is incompatible with the necessary condition for the existence of a zero mode, i.e., $m=[1+{\rm sgn}(Jf)]/2$. 
We therefore conclude that there does exist one zero mode, which is also the \textit{exact} zero mode, for large odd $L$. It might be intriguing that just one zero mode appears regardless of the parity of $L$ for large chains, in spite of the fact that the exact zero mode exists only when $L$ is odd.

\subsection{Majorana zero operators in inhomogeneous chains with $\mu=0$}
\label{sec:mu0 zero modes inhomogeneous}

The Majorana zero operators localized at the boundary exist even in the presence of couplings varying over space. To see this, we consider the inhomogeneous chain with $\phi_{1} = \phi_{2} = \pi/2$ and $\mu=0$. The Hamiltonian in terms of $a_j$ and $b_j$ reads
\begin{equation}
H = \frac{\ii}{2} \left[ \sum_{j=1}^{L-1} \left( J_{j} b_{j} a_{j+1} - f_{j} a_{j} b_{j+1} \right) + b \left( f_{L} a_{1} a_{L} + J_{L} b_{1} b_{L} \right) \right].
\end{equation}
We here assume that the minimum of $\left\{ J_{j} \right\}$ is larger than the maximum of $\left\{ f_{j} \right\}$ : $\min_{j} \left| J_{j} \right| > \max_{j} \left| f_{j} \right|$. Then, when we set $K := \lfloor \left( L+1 \right) /2 \rfloor$, the Majorana zero operators take the following forms:
\onecolumngrid
\begin{eqnarray}
\Psi_{1} &=& a_{1} + \sum_{m=1}^{K-1} \left( - \frac{f_{1}}{J_{2}} \right) \cdots \left( - \frac{f_{2m-1}}{J_{2m}} \right) a_{2m+1}
+ {\sf b}\, \sum_{m=1}^{K-1} \left( - \frac{f_{L}}{J_{L-1}} \right) \cdots \left( - \frac{f_{L-2m+2}}{J_{L-2m+1}} \right) b_{L-2m+1},  
	\label{eq: zero mode1 mu0 inhomo} \\
\Psi_{2} &=& b_{L} - \frac{{\sf b} J_{L}}{J_{1}} a_{2}
+ \sum_{m=1}^{K-1} \left( - \frac{f_{L-1}}{J_{L-2}} \right) \cdots \left( - \frac{f_{L-2m+1}}{J_{L-2m}} \right) b_{L-2m}
- \frac{{\sf b} J_{L}}{J_{1}} \sum_{m=1}^{K-2} \left( - \frac{f_{2}}{J_{3}} \right) \cdots \left( - \frac{f_{2m}}{J_{2m+1}} \right) a_{2m+2}.
	\label{eq: zero mode2 mu0 inhomo}
\end{eqnarray}
\twocolumngrid
These zero operators remain normalizable even in the $L \to \infty$ limit. For the homogeneous chain with odd $L$, Eqs.~(\ref{eq: zero mode1 mu0 inhomo}) and (\ref{eq: zero mode2 mu0 inhomo}) reduce to Eqs.~(\ref{mu0 zero mode1}) and (\ref{mu0 zero mode2}), respectively. When the amplitude on the boundaries vanishes (${\sf b} = 0$), the above zero modes boil down to those of the open chain.

In the case of odd $L$, the Majorana zero operators exactly commute with the Hamiltonian:
\begin{equation}
\left[ H, \Psi_{1} \right]
= \left[ H, \Psi_{2} \right]
=0
\end{equation}
On the other hand, in the case of even $L$, the zero operators do not exactly commute with the Hamiltonian. However, their commutators with $H$ are exponentially small in the system size:
\begin{eqnarray}
\left[ H, \Psi_{1} \right] &=& -{\ii} \left[ \prod_{m=1}^{K} \left( - \frac{f_{2m-1}}{J_{2m}} \right) \right] J_{L} b_{L} \nonumber \\
&&~~~~+{\ii}\, {\sf b} \left[ \prod_{m=1}^{K} \left( - \frac{f_{2m}}{J_{2m-1}} \right) \right] J_{1} a_{2},
\end{eqnarray}
\begin{eqnarray}
\left[ H, \Psi_{2} \right]\!\! &=&\!\! -{\ii} \left[ \prod_{m=1}^{K-1} \left( - \frac{f_{2m+1}}{J_{2m}} \right) \right] f_{1} a_{1} \nonumber \\
&&+{\ii}\, \frac{{\sf b}\, J_{L}}{J_{1}} \left[ \prod_{m=1}^{K-1} \left( - \frac{f_{2m}}{J_{2m+1}} \right) \right] J_{L-1} b_{L-1},
\end{eqnarray}
implying that each commutes with $H$ in the limit $L \to \infty$. Therefore, the Majorana zero operators do exist in the $L \to \infty$ limit, regardless of the parity of $L$. This clearly demonstrates that the topological order survives even in the presence of disorder.

\section{Effects of nearest-neighbor interactions}
\label{sec: interaction}

\begin{figure}[b]
	\includegraphics[width=0.95\columnwidth]{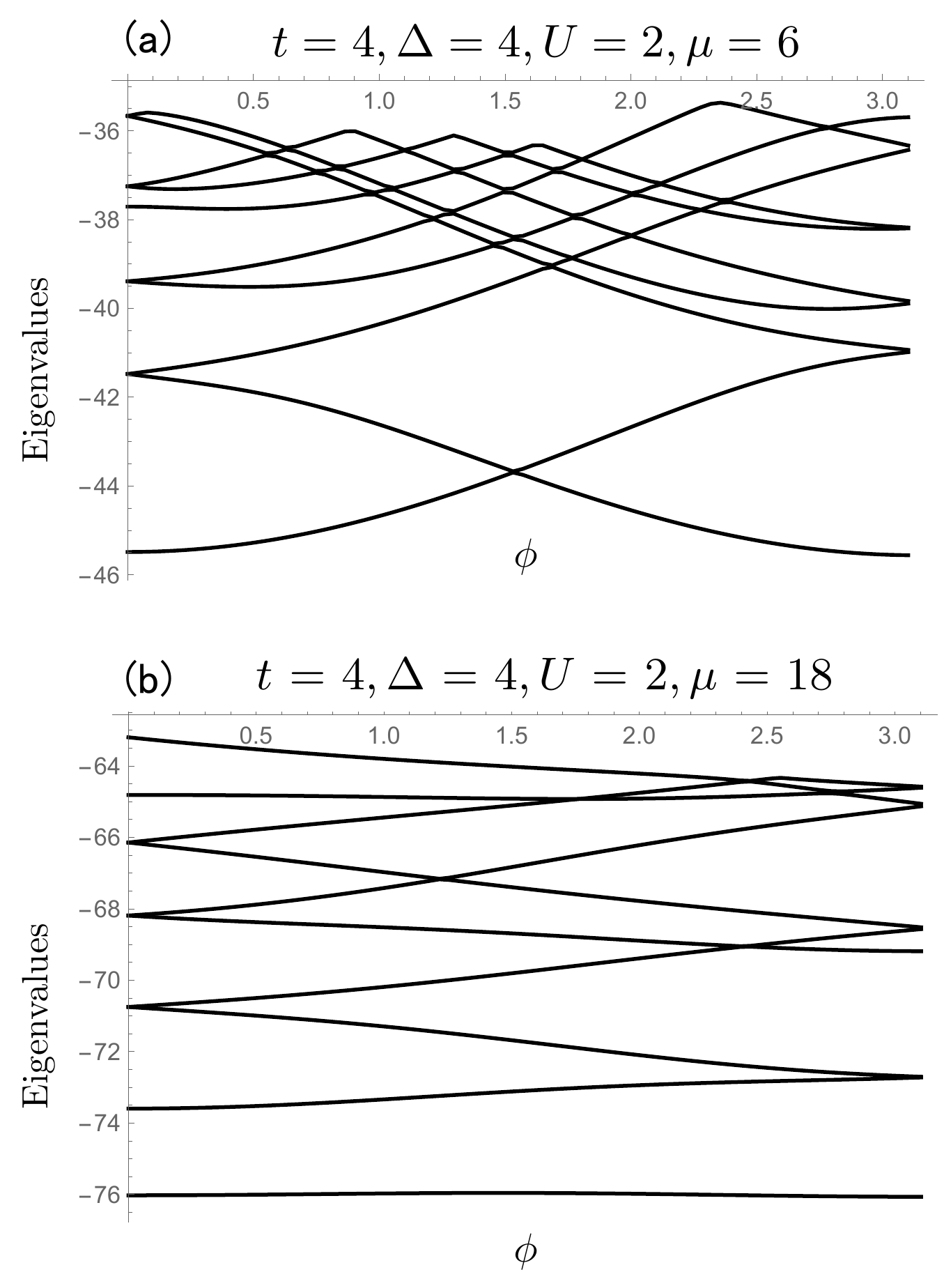} 
	\caption{Spectrum of the interacting Hamiltonian $H_{\rm{int}}|_{{\sf a}={\sf b}=1;(\phi, \phi)}$ ($0\le\phi\le\pi$), for the topological phase and the trivial phase. Calculations are performed for $L=16$, and the lowest ten eigen-energies are shown. (a) topological phase ($t=4, \Delta=4, U=2, \mu=6$): the level crossing between the two lowest-lying states occurs at $\phi \simeq \pi/2$. (b) trivial phase ($t=4, \Delta=4, U=2, \mu=18$): the spectral gap above the ground state never closes at any $\phi$.}
      \label{fig:int}
\end{figure}

In this section, we consider the generalization of $H$ to the case with nearest-neighbor interactions~\cite{Hassler-12}:
\begin{eqnarray}
H_{\rm{int}}=H + \sum^{L}_{j=1} U \left( 2n_{j}-1 \right) \left( 2n_{j+1}-1 \right),
\label{eq: interactingH}
\end{eqnarray}
where $U$ is the strength of the interaction and $n_j:=c^\dag_jc_j$ is the fermion number operator at site $j$. For the non-interacting Kitaev chains, we have demonstrated that the ground-state fermionic parity changes between the PBC and the APBC, when the system resides in the topological phase. Now we remark that a small $U$ that fails to close the many-body gap will not remove the parity switch, which leads to the same conclusion as the free case. To illustrate 
this heuristic picture, we have performed exact diagonalization of interacting chains. Figure~\ref{fig:int} shows the evolution of the spectrum as a 
function of the phase parameter $\phi := \phi_1 = \phi_2$. In the topological phase, the level crossing between the two lowest-lying states occurs (see Fig.~\ref{fig:int} (a)). On the other hand, this does not happen in the trivial phase (see Fig.~\ref{fig:int} (b)). These results imply that the level crossing is a generic feature of topological phases even in the presence of nearest-neighbor interactions. Besides, the parity switches indeed take place for the exactly solvable interacting chains with the fine-tuned parameters satisfying `frustration-free condition\cite{Katsura_Int_Majorana}' (see Appendix \ref{sec:frustration-free}). 

\begin{figure}[t]
	\includegraphics[width=0.93\columnwidth]{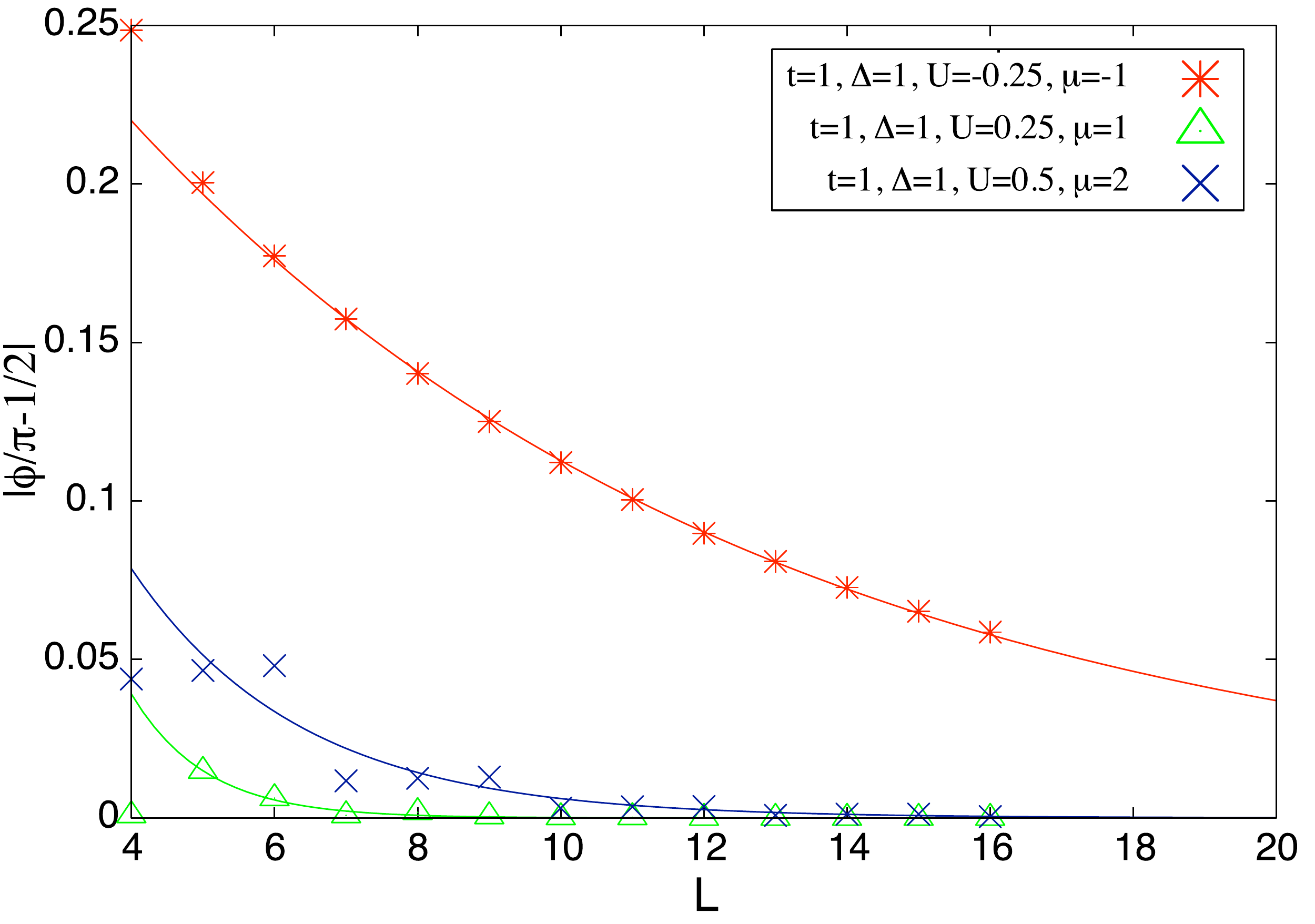}
   \includegraphics[width=0.93\columnwidth]{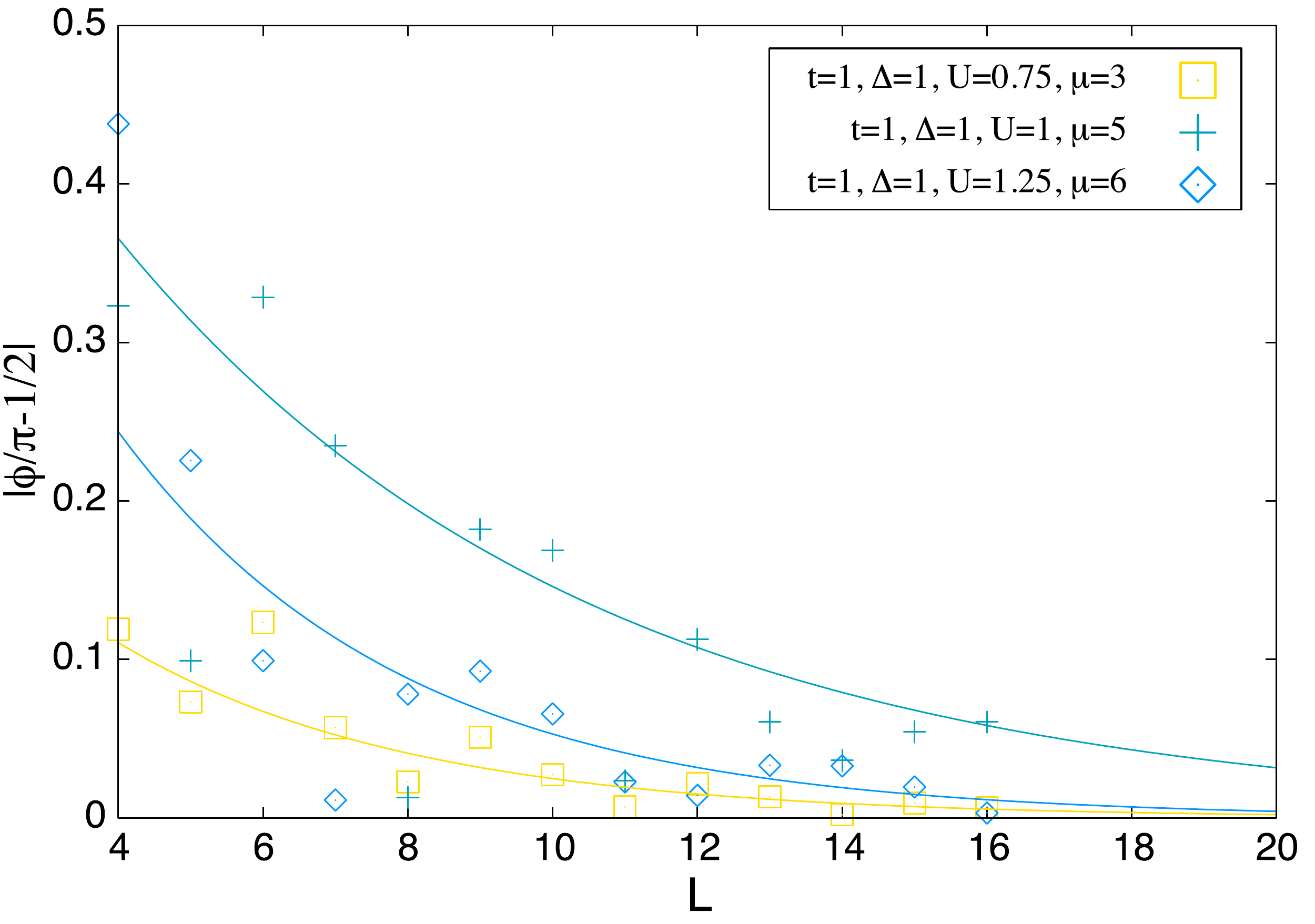} 
	\caption{Dependence of the level crossing angle $\lvert\phi/\pi-1/2\rvert$ on the system size. 
Fitting functions are shown by solid curves, which demonstrate the exponential decay of $\lvert\phi/\pi-1/2\rvert$ with increasing $L$. The values of the parameters $t, \Delta, U, \mu$ are taken within the region of the topological phase.}
      \label{fig:degangle}
\end{figure}

For free chains, we have also explicitly determined the parameter conditions for the two-fold degeneracy of the ground states. In particular, the chains with $\phi=\pi/2$ or $3\pi/2$ definitely satisfy the obtained conditions in the infinite-size limit. Now let us see what 
happens in interacting systems. We can observe the level crossing appears at around $\phi=\pi/2$ in Fig.~\ref{fig:int} (a), hence it may be conjectured that the two-fold degeneracy of the ground states at $\phi=\pi/2$ is also the case with interacting chains in the infinite-size limit, irrespective of the details of the other parameters. We have performed exact diagonalization further to confirm the conjecture. Figure~\ref{fig:degangle} provides the difference between the level crossing angle $\phi$ and $\pi/2$ as a function of the system size $L$. Remarkably, $\lvert\phi-\pi/2\rvert$ shows a significant decrease as $L$ is increased. As discussed in Sec. \ref{sec:emergence of zero modes}, $\lvert\phi-\pi/2\rvert$ has exponentially small ($\sim\varepsilon(L)$ in Eq. (\ref{eq:eps_L}) ) dependence on the system size $L$ in free chains. 
Fitting the data with an exponential ansatz, $|\phi-\pi/2| \propto \exp (-L/\xi )$, yields the curves shown in Fig.~\ref{fig:degangle}. 
%
%
The exponential decrease of $\lvert\phi-\pi/2\rvert$ in $L$ clearly demonstrates that the ground states are two-fold degenerate at $\phi=\pi/2$ in the infinite-size limit even for interacting chain. In addition, the ground-state degeneracy at the crossing point suggests the presence of zero modes at $\phi=\pi/2$. An interesting question is whether the zero modes that map one of the ground state to the other are strong or weak. This could be studied systematically following previous approaches~\cite{Stoudenmire-11, Aris-15}, but we leave it for future work. 

\section{Conclusion}
\label{sec:conc}

In this paper, we have studied the Kitaev chains under generalized twisted boundary conditions characterized by the phase parameters $(\phi_1, \phi_2)$. We found that the phases $(\phi_1, \phi_2)$ can be adjusted so that Majorana zero modes appear as long as the bulk couplings are those of the Kitaev chain in the topological phase. By computing the Pfaffian of the Hamiltonian matrix in the Majorana basis, we rigorously obtained the condition on $(\phi_1, \phi_2)$ for the presence of Majorana zero modes. The condition reduces to $\phi_1=\phi_2=\pi/2$ or $3\pi/2$ in the infinite-size limit when the constraint $\phi_1=\phi_2$ is imposed. 

We then analyzed finite chains at $\phi_1=\phi_2=\pi/2$ and enumerated conditions on the other parameters under which exact Majorana zero modes exist. A particularly interesting case is $\mu=0$, where the exact zero modes must appear in a chain of odd length, irrespective of the details of the other parameters. The full energy spectrum for this case was analyzed in detail and the explicit expressions for the Majorana zero operators that commute with the Hamiltonian were obtained. The operators obtained are exponentially localized and hence normalizable in the infinite-size limit. We also showed that the presence of Majorana zero operators survive even in the presence of spatially varying couplings, provided that $\phi_1=\phi_2=\pi/2$ and $\mu=0$. 

The robustness of the zero modes at least in the weak sense persists even in the presence of interactions, as demonstrated by our analytical and numerical results. Whether the zero modes at the level crossing point are strong or weak is an intriguing open question. It would also be interesting to see if and how the twisted boundary conditions lead to a level crossing in other systems such as parafermion~\cite{Fendley12, Clarke-14, KlinovajaLoss14, Mong-14, Jermyn-14, Aris-15, Alicea-16} and XYZ chains~\cite{Fendley-XYZ}, which are not reducible to free fermions.

\section*{Acknowledgment}
The authors thank Yutaka Akagi and Ken Shiozaki for valuable discussions, and Domenico Giuliano for bringing our attention to Ref.~\onlinecite{Giuliano-16}. 
H. K. was supported in part by JSPS KAKENHI Grant No. JP15K17719, No. JP16H00985, and No. JP15K21717. N. W. was supported by a startup fund from the Beijing Institute of Technology.

\appendix
\section{Diagonalization of the Kitaev chain for the PBC and the APBC}
\label{APP_PAP}

The Hamiltonian $H$ with the PBC or the APBC can be easily diagonalized by the Fourier transform. Let $\psi_k$ be the Fourier transform of $c_j$. The annihilation operator $c_j$ is written in terms of $\psi_k$ as
\begin{align}
c_{j} = \frac{e^{ - \ii \pi/4}}{\sqrt{L}} \sum_{k \in \mathcal{K}} e^{\ii kj} \psi_{k},
	\label{eq:Fermi Fourier}
\end{align}
where $\mathcal{K}$ denotes the set of possible wavenumbers which depends on both the parity of $L$ and the boundary condition. 
Then the Hamiltonian in $k$-space is given by
\begin{eqnarray}\label{FourierHam}
H \!\!&=&\!\! \sum_{k\in\mathcal{K} \atop k\neq 0,\pi}H_{k} + 
\left\{
\begin{array}{cl}
H_0 + H_\pi & (L\!=\!{\rm even}, {\rm PBC}) \\
0 & (L\!=\!{\rm even}, {\rm APBC}) \\
H_0 & (L\!=\!{\rm odd}, {\rm PBC}) \\
H_\pi & (L\!=\!{\rm odd}, {\rm APBC})
\end{array}
\right.\!\!\!\!,
\end{eqnarray}
where
\begin{eqnarray}
H_{k} \!=\! - \begin{pmatrix} \psi_{k}^{\dagger} & \psi_{-k} \end{pmatrix} \!\!
\begin{pmatrix} t \cos k + \mu/2 & \Delta \sin k \\ \Delta \sin k & -t \cos k - \mu/2  \end{pmatrix}\!\!
\begin{pmatrix} \psi_k \\ \psi_{-k}^\dagger \end{pmatrix}. \nonumber
\\
\end{eqnarray}
For $k\neq 0, \pi$, the ground state of $H_{k}+H_{-k}$ is the vacuum of Bogoliubov quasiparticles:
\begin{eqnarray}
\lvert\mathrm{g.s.}\rangle_k = \alpha_k\alpha_{-k}\lvert0\rangle_k,
\end{eqnarray}
where $\lvert0\rangle_k$ is the vacuum state of $\psi_k, \psi_{-k}$, and $\alpha_{k/-k}$ is the annihilation operator of Bogoliubov quasiparticles, which has odd fermionic parity. Therefore, the ground state at each $k$ has even number of fermions unless $k=0$ or $\pi$. 
For $k=0, \pi$, the fermionic parity in the ground state of $H_k$ is given by
\begin{enumerate}
\item $H_0$
\begin{eqnarray}
P\mid_{k=0}= \begin{cases} 1 & 2t+\mu<0 \\ -1 & 2t+\mu>0\end{cases},
\end{eqnarray}
\item $H_{\pi}$
\begin{eqnarray}
P\mid_{k=\pi}= \begin{cases} 1 & -2t+\mu<0 \\ -1 & -2t+\mu>0\end{cases}.
\end{eqnarray}
\end{enumerate}
Now, it is easy to get the fermionic parity in the ground state and hence Eq.~(\ref{eq:SpecIntersection}).

\section{The specific form of the coefficient matrix $X$ in Eq.~(\ref{zero mode matrix})}
\label{sec:zero mode matrix}

The coefficient matrix $X$ in Eq.~(\ref{zero mode matrix}) which determines whether zero modes exist is
\onecolumngrid
\begin{equation}
X = \left( \begin{array}{@{\,}cccc@{\,}}
{\sf b} \left( t- \Delta \right) \lambda_{+}^{L}
& {\sf b} \left( t- \Delta \right) \lambda_{-}^{L}
& \left( \tilde{\sf a} \mu + \left( t + \Delta \right) \lambda_{+} \right) \lambda_{+}^{L}
& \left( \tilde{\sf a} \mu + \left( t + \Delta \right) \lambda_{-} \right) \lambda_{-}^{L} \\
- \tilde{\sf a} \mu \lambda_{+} - \left( t - \Delta \right)
& - \tilde{\sf a} \mu \lambda_{-} - \left( t - \Delta \right)
& {\sf b} \left( t+ \Delta \right) \lambda_{+}
& {\sf b} \left( t+ \Delta \right) \lambda_{-} \\
{\sf b} \left( t - \Delta \right) \lambda_{+}
& {\sf b} \left( t - \Delta \right) \lambda_{-}
& - \tilde{\sf a} \mu \lambda_{+} - \left( t - \Delta \right)
& - \tilde{\sf a} \mu \lambda_{-} - \left( t - \Delta \right) \\
\left( \tilde{\sf a} \mu + \left( t + \Delta \right) \lambda_{+} \right) \lambda_{+}^{L}
& \left( \tilde{\sf a} \mu + \left( t + \Delta \right) \lambda_{-} \right) \lambda_{-}^{L}
& {\sf b} \left( t + \Delta \right) \lambda_{+}^{L}
& {\sf b} \left( t + \Delta \right) \lambda_{-}^{L} \\		
      	\end{array} \right).
\end{equation}
In the case of OBC ($\tilde{\sf a} = {\sf b} = 0$), $X$ becomes block diagonal and
\begin{eqnarray}
\det X
&=& \det \left( \begin{array}{@{\,}cc@{\,}}
 \left( t - \Delta \right)
& \left( t - \Delta \right) \\
\left( t + \Delta \right) \lambda_{+}^{L+1}
& \left( t + \Delta \right) \lambda_{-}^{L+1} \\
			\end{array} \right)
\times \det \left( \begin{array}{@{\,}cc@{\,}}
\left( t + \Delta \right) \lambda_{+}^{L+1}
& \left( t + \Delta \right) \lambda_{-}^{L+1} \\
\left( t - \Delta \right)
& \left( t - \Delta \right) \\
			\end{array} \right)
\propto \left( \lambda_{+}^{L+1} - \lambda_{-}^{L+1} \right)^{2}.
\end{eqnarray}
\twocolumngrid

\section{The detailed calculation of the full spectrum of the chain with $\mu=0$}
\label{sec:mu0 detail}

In this Appendix, we present a detailed exposition of the calculation of the full spectrum of the chain with $\mu=0$. The results depend on the parity of $L$, but the way to get the spectrum is essentially the same. We first conduct a plane-wave expansion in the bulk, and then match the solutions at the boundaries. The calculation of the chain with odd $L$ is easier than that of the chain with even $L$: we have to evaluate the determinant of a $2 \times 2$ matrix in the case of odd $L$, but on the other hand, we have to analyze a $4 \times 4$ matrix in the case of even $L$.

\subsection{Even $L$}
\label{sec:}

The spectrum of the original $2L \times 2L$ Hamiltonian is equivalent to that of the following $2L \times 2L$ matrix:
\begin{equation}
H^{\left( \mathrm{even} \right)} = \frac{\ii}{4} \left( \begin{array}{@{\,}ccccccccc@{\,}}
								 0 & -f & & & &  & & & {\sf b} f \\
								 f & 0 & J & & &  & & & \\
								  & -J & 0 & -f & &  & & & \\
								  & & \ddots & \ddots & \ddots & & & & \\
								  & & & f & 0 & - {\sf b} J & & & \\
								  & & & & {\sf b} J & 0 & J & & \\
								  & & & & & \ddots & \ddots & \ddots & \\
								  & & & & & & f & 0 & J \\
								 - {\sf b} f & & & & & & & -J & 0 \\								
								       						\end{array} \right).
\end{equation}								       						
Let $\left( C_{1},~D_{1},~\cdots, C_{M},~D_{M},~E_{1},~F_{1},~\cdots, E_{M},~F_{M}\right)$ be an eigenvector of $H^{\left( \mathrm{even} \right)}$, where $M:=L/2$. The relation between $\{ A_{j} \},~\{ B_{j} \}$ and $\{ C_{j} \},~\{ D_{j} \},~\{ E_{j} \},~\{ F_{j} \}$ is that
\begin{equation}
A_{2j-1} = C_{j},~A_{2j} = F_{j},~B_{2j-1} = E_{j},~B_{2j} = D_{j},
\end{equation}
where $j=1,2, \cdots, M$. If we take an ansatz $C_{j} \sim c e^{\ii qj},~D_{j} \sim d e^{\ii qj}$, the bulk conditions give
\begin{equation}
\left( \begin{array}{@{\,}cc@{\,}}
	\ii \left( Je^{\ii q} + f \right) & - \varepsilon \\
	\varepsilon & \ii \left( Je^{-\ii q} + f \right) \\
	\end{array} \right) \left( \begin{array}{@{\,}cc@{\,}}  c \\ d \\ \end{array} \right) = 0,
\end{equation}
and the determinant of the coefficient matrix should be zero for the existence of the eigenvectors, which is just the dispersion relation
\begin{equation}
\varepsilon = \pm \left| Je^{iq} + f \right| = \pm \sqrt{J^{2} + f^{2} + 2Jf \cos q}.
\end{equation}
Remembering that we focus on the non-negative eigenvalues ($E, \varepsilon \geq 0$),
\begin{equation}
\frac{d}{c} = \ii \frac{Je^{\ii q} + f}{\varepsilon} = \ii \sqrt{\frac{Je^{\ii q} + f}{Je^{-\ii q} + f}} =: C \left( q \right).
\end{equation}
Therefore, we can expand $\{ C_{j} \},~\{ D_{j} \}$ as follows
\begin{eqnarray}
C_{j} &=& c_{+} e^{\ii qj} + c_{-} e^{-\ii qj}, \nonumber \\
D_{j} &=& c_{+} C \left( q \right) e^{\ii qj} - \frac{c_{-}}{C \left( q \right)} e^{-\ii qj}.
\end{eqnarray}
We can expand $\{ E_{j} \},~\{ F_{j} \}$ in a similar way:
\begin{eqnarray}
E_{j} &=& e_{+} e^{\ii qj} + e_{-} e^{- \ii qj}, \nonumber \\
F_{j} &=& \frac{e_{+}}{C \left( q \right)} e^{\ii q (j+1)} - e_{-} C \left( q \right) e^{- \ii q (j+1)}.
\end{eqnarray}
Replacing $\{ A_{j} \}$,~$\{ B_{j} \}$ in Eq.~(\ref{mu0 boundary conditions}) with $\{ C_{j} \}$,~$\{ D_{j} \}$,~$\{ E_{j} \}$,~$\{ F_{j} \}$, we obtain the boundary conditions:
\begin{eqnarray}
JD_{0} + {\sf b} fF_{M} &=& 0 \nonumber \\
{\sf b} E_{1} + C_{M+1} &=& 0 \nonumber \\
fF_{0} - {\sf b} JD_{M} &=& 0 \nonumber \\
{\sf b} C_{1} - E_{M+1} &=& 0.
\end{eqnarray}
Substituting the plane wave expansions into the boundary conditions, the consistency condition for the wavenumber $q$ is obtained as
\begin{equation}
X \left( \begin{array}{@{\,}cccc@{\,}}  c_{+} & c_{-} & e_{+} & e_{-} \\ \end{array} \right)^{T} = 0,
\end{equation}
where $X$ is the following $4 \times 4$ matrix:
\begin{equation}
\left( \begin{array}{@{\,}cccc@{\,}}
		{\sf b} e^{\ii q} & {\sf b} e^{-\ii q} & - e^{\ii q (M+1)} & - e^{-\ii q (M+1)} \\
		e^{\ii q (M+1)} & e^{-\ii q (M+1)} & {\sf b} e^{\ii q} & {\sf b} e^{-\ii q} \\
		J C^{2} & -J & {\sf b} f e^{\ii q (M+1)} & -{\sf b} f C^{2} e^{-\ii q (M+1)} \\
		{\sf b} J C^{2} e^{\ii qM} & -{\sf b} J e^{-\ii qM} & -f e^{\ii q} & f C^{2} e^{-\ii q} \\
	\end{array} \right).
\end{equation}
The above equation has a nontrivial solution if the determinant of the coefficient matrix vanishes. After some calculations, this condition boils down to Eq.~(\ref{quantization condition for even}).

\subsection{Odd $L$}
\label{sec:}

The spectrum of the original $2L \times 2L$ Hamiltonian is equivalent to those of the two independent $L \times L$ matrices, one of which is
\begin{equation}
H^{\left( \mathrm{odd} \right)} = \frac{\ii}{4} \left( \begin{array}{@{\,}cccccc@{\,}}
	0 & -f & & & & {\sf b} f \\
	f & 0 & J & & & \\
	& -J & 0 & -f & & \\
	& & \ddots & \ddots & \ddots & \\
	& & & f & 0 & J \\
	- {\sf b} f & & & & -J & 0 \\ \end{array} \right).
\end{equation}
We restrict our attention to the spectrum of this matrix since that of the other one is obtained by exchanging $(J,f)$ for $(-f,-J)$ in this one. Let $\left( C_{1},~D_{1},~\cdots, C_{N-1},~D_{N-1},~C_{N} \right)$ be an eigenvector of $H^{\left( \mathrm{odd} \right)}$, where $N := (L+1)/2$. The relation between $\{ A_{j} \},~\{ B_{j} \}$ and $\{ C_{j} \},~\{ D_{j} \}$ is that
\begin{equation}
A_{2j-1} = C_{j},~B_{2j} = D_{j},~A_{L} = C_{N},
\end{equation}
where $j=1,2, \cdots, N-1$. The bulk conditions are the same as those of the case of even $L$, and the dispersion relation is thus the same:
\begin{equation}
\varepsilon = \pm \sqrt{J^{2} + f^{2} + 2Jf \cos q}
\end{equation}
We can expand the eigenstates as
\begin{eqnarray}
C_{j} &=& c_{+} e^{\ii qj} + c_{-} e^{-\ii qj}, \nonumber \\
D_{j} &=& c_{+} C \left( q \right) e^{\ii qj} - \frac{c_{-}}{C \left( q \right)} e^{-\ii qj}.
\end{eqnarray}
The boundary conditions in terms of $\{ C_{j} \},~\{ D_{j} \}$ are
\begin{equation}
JD_{0} + {\sf b} f C_{N} = 0,~{\sf b} C_{1} - D_{N} = 0.
\end{equation}
Substituting the plane wave expansions into the boundary conditions, the consistency condition for the wavenumber $q$ is obtained as
\begin{equation}
\left( \begin{array}{@{\,}cc@{\,}}
	J C^{2} + {\sf b} f C e^{\ii qN} & - J + {\sf b} f C e^{-\ii qN} \\
	{\sf b} C e^{\ii q} - C^{2} e^{\ii qN} & {\sf b} C e^{-\ii q} + e^{-\ii qN} \\
	\end{array} \right) \left( \begin{array}{@{\,}cc@{\,}}  c_{+} \\ c_{-} \\ \end{array} \right) = 0.
\end{equation}
This equation has a nontrivial solution if the determinant of the coefficient matrix vanishes. This condition reads
\begin{equation}
Je^{\ii q} + f = 0~~\mathrm{or}~~\frac{\sin qN}{\sin q \left( N-1 \right)} = \frac{{\sf b}^{2}f}{J}.
\end{equation}
``$Je^{\ii q} + f = 0$'' means the existence of the exact zero mode. And the condition for the other Hamiltonian is obtained by swapping $( J,f )$ for $( -f, -J )$ in the above condition, which leads to the quantization condition Eq.~(\ref{quantization condition for odd}).

\section{Parity switches of the ground states for the interacting chains satisfying the frustration-free condition}
\label{sec:frustration-free}

It was shown for ${\sf a}=1/2$ and ${\sf b}=0$ that the exact ground states of $H_{\rm{int}}|_{{\sf a}=1/2, {\sf b}=0}$ can be obtained, provided that the parameters satisfy the following `frustration-free' condition~\cite{Katsura_Int_Majorana}
\begin{eqnarray}
\mu = \mu^{*} := 4\sqrt{U^2+tU+\frac{t^2-\Delta^2}{4}},
	\label{frus_free}
\end{eqnarray}
The ground states are found to be two-fold degenerate and have opposite fermionic parities
\begin{eqnarray}
\lvert\Psi^{\rm{(even/odd)}}_0\rangle=\frac{1}{(1+\alpha^2)^{L/2}}A_L^{\rm{(even/odd)}}\lvert\mathrm{vac}\rangle,
\end{eqnarray}
where $\alpha := \sqrt{\cot \left[ \arctan \left( 2\Delta/\mu^{*} \right)/2 \right]}$, $A_{L}^{\rm{(even/odd)}} := A_{L}^{(+)} \pm A_{L}^{(-)}$ and
\begin{eqnarray}
A_{L}^{(\pm)} := 
e^{\pm\alpha c^{\dag}_{1}} e^{\pm\alpha c^{\dag}_{2}} \cdots  e^{\pm\alpha c^{\dag}_{L}}.
\end{eqnarray}
Here, $|\rm{vac}\rangle$ is the vacuum state of $\{c_j\}$. 
The key to obtaining the exact ground states is the fact that the two-site states $e^{\pm\alpha c^\dag_j}e^{\pm\alpha c^\dag_{j+1}}|\rm{vac}\rangle$ 
minimize the following local Hamiltonian $h_{j}$ simultaneously 
\begin{eqnarray}
h_{j} &=&
-t \left( c^{\dag}_{j} c_{j+1} + \mathrm{h.c.} \right) + \Delta \left( c_{j} c_{j+1} + \mathrm{h.c.} \right) \nonumber\\
&-& \frac{\mu^{*}}{2} \left( n_{j} + n_{j+1}-1 \right) + U \left( 2n_{j}-1 \right) \left( 2n_{j+1}-1 \right). \nonumber \\
\end{eqnarray}
In terms of $h_j$ ($j=1,2,\cdots L-1$) the total Hamiltonian is expressed as $H_{\rm int}=\sum^{L-1}_{j=1} h_j$. We note in passing that the phase diagram of the model has been obtained in Ref.~\onlinecite{Hassler-12} and the frustration-free line always resides in the topological phase~\cite{Katsura_Int_Majorana}. 

One can show that, under the frustration-free condition Eq.~(\ref{frus_free}), $|\Psi^{\rm{(odd)}}_0\rangle$  ($|\Psi^{\rm{(even)}}_0\rangle$) is the ground state of the interacting Kitaev chain under the PBC (APBC). We note that the same result has also been presented~\cite{Shiozaki, Shiozaki2}, but for the reader's convenience we give an explicit proof of this fact. Since $|\Psi^{\rm{(odd)}}_0\rangle$  and $|\Psi^{\rm{(even)}}_0\rangle$ are the ground states of the bulk part of the Hamiltonian, it suffices to show that they are ground states of the boundary part $h_L$. The proof goes as follows:
\begin{enumerate}
\item{PBC: $c_{L+1}=c_{1}, c^{\dagger}_{L+1}=c^{\dagger}_1$.} 
\newline
We first note that any state of the form $e^{\pm\alpha c^\dag_L}e^{\pm\alpha c^\dag_{1}} \cdots |\rm{vac}\rangle$, is a ground states of $h_L$, where the part $\cdots$ denotes an arbitrary polynomial in $c^\dag_2, ..., c^\dag_{L-1}$. Consequently, any linear combination of such states is also a ground state of $h_L$. Then we rewrite $A_L^{(\rm{odd})}$ as
\begin{equation}
\begin{split}
A_L^{(\rm{odd})} &= A_{L-1}^{(\rm{even})}e^{\alpha c_L^\dagger}-2A_{L-1}^{(-)} \\
&= e^{\alpha c_L^\dagger}A_{L-1}^{(\rm{even})}-2A_{L-1}^{(-)} \\
&= e^{\alpha c^\dag_L}e^{\alpha c^\dag_1} \cdots  e^{\alpha c^\dag_{L-1}}-e^{-\alpha c^\dag_L}e^{-\alpha c^\dag_1} \cdots  e^{-\alpha c^\dag_{L-1}},
\end{split}
\end{equation}
and find that $\lvert\Psi^{\rm{(odd)}}_0\rangle\propto A_L^{\rm{(odd)}}\lvert\mathrm{vac}\rangle$ minimizes $h_L$. This is the unique ground state of $H_{\rm int}$ at $\mu=\mu^*$ with PBC,  because the other state $\lvert\Psi^{\rm{(even)}}_0\rangle$ does not minimizes $h_L$. 

\item{APBC: $c_{L+1}=-c_1, c^\dagger_{L+1}=-c^\dagger_1$.} 
\newline
We first note that any state of the form $e^{\pm\alpha c^\dag_L}e^{\mp \alpha c^\dag_{1}} \cdots |\rm{vac}\rangle$, is a ground states of $h_L$, where the part $\cdots$ denotes an arbitrary polynomial in $c^\dag_2, ..., c^\dag_{L-1}$. Consequently, any linear combination of such states is also a ground state of $h_L$. Then we rewrite $A_L^{(\rm{even})}$ as
\begin{equation}
\begin{split}
A_L^{(\rm{even})} &= A_{L-1}^{(\rm{odd})}e^{\alpha c_L^\dagger}+2A_{L-1}^{(-)} \\
&= e^{-\alpha c_L^\dagger}A_{L-1}^{(\rm{odd})}+2A_{L-1}^{(-)} \\
&= e^{-\alpha c^\dag_L}e^{\alpha c^\dag_1} \cdots  e^{\alpha c^\dag_{L-1}}+e^{\alpha c^\dag_L}e^{-\alpha c^\dag_1} \cdots  e^{-\alpha c^\dag_{L-1}}.
\end{split}
\end{equation}
and find that $\lvert\Psi^{\rm{(even)}}_0\rangle\propto A_L^{\rm{(even)}}\lvert\mathrm{vac}\rangle$ minimizes $h_L$. This is the unique ground state of $H_{\rm int}$ at $\mu=\mu^*$ with APBC,  because the other state $\lvert\Psi^{\rm{(odd)}}_0\rangle$ does not minimizes $h_L$. 
\end{enumerate}

The change of the parity between the PBC and the APBC for the interacting Kitaev model again indicates the ground-state degeneracy at some point on a path $(\phi_1(s),\phi_2(s))$ that connects $H_{\rm{int}}|_{{\sf a}={\sf b}=1;(0,0)}$ and $H_{\rm{int}}|_{{\sf a}={\sf b}=1;(\pi,\pi)}$. Therefore, the level crossing in the spectrum survives in the presence of interactions, under the frustration-free condition.



\begin{thebibliography}{}

\bibitem{Wilczek09}
F. Wilczek, Nature Phys. {\bf 5},  614  (2009).

\bibitem{Alicea12}
J. Alicea, Rep. Prog. Phys. {\bf 75},  076501  (2012).

\bibitem{LeijnseFlensberg12}
M. Leijnse and K. Flensberg, Semicond. Sci. Technol. {\bf 27},  124003  (2012).

\bibitem{ElliottFranz15}
S. R. Elliott and M. Franz, Rev. Mod. Phys. {\bf 87}, 137 (2015).

\bibitem{Mourik-12}
V. Mourik, K. Zuo, S. M. Frolov, S. R. Plissard, E. P. A. M. Bakkers, and L. P. Kouwenhoven, Science {\bf 336},  1003  (2012).

\bibitem{Rokhinson-12}
L. P. Rokhinson, X. Liu, and J. K. Furdyna, Nature Phys. {\bf 8}, 795  (2012).

\bibitem{Das-12}
A. Das, Y. Ronen, Y. Most, Y. Oreg, M. Heiblum, and H. Shtrikman, Nature Phys. {\bf 8},  887  (2012).

\bibitem{DengYu-12}
M. T. Deng, C. L. Yu, G. Y. Huang, M. Larsson, P. Caroff, and H. Q. Xu, Nano Letters {\bf 12},  6414  (2012).

\bibitem{Finck-13}
A. D. K. Finck, D. J. Van Harlingen, P. K. Mohseni, K. Jung, and X. Li, Phys. Rev. Lett. {\bf 110}, 126406 (2013).

\bibitem{Churchill-13}
H. O. H. Churchill, V. Fatemi, K. Grove-Rasmussen, M. T. Deng, P. Caroff, H. Q. Xu, and C. M. Marcus, Phys. Rev. B {\bf 87},  241401  (2013).

\bibitem{Nadj-Perge-14}
S. Nadj-Perge, I. K. Drozdov, J. Li, H. Chen, S. Jeon, J. Seo, A. H. MacDonald, B. A. Bernevig, and A. Yazdani, Science {\bf 346},  602  (2014).

\bibitem{Higginbotham-15}
A. P. Higginbotham, S. M. Albrecht, G. Kir\v{s}anskas, W. Chang, F. Kuemmeth, P. Krogstrup, T. S. Jespersen, J. Nyg\r{a}rd, K. Flensberg,and C. M. Marcus, Nat. Phys. {\bf 11}, 1017 (2015).

\bibitem{Ivanov-01}
D. A. Ivanov, Phys. Rev. Lett. {\bf 86}, 268 (2001).

\bibitem{Nayak-08}
C. Nayak, S. Simon, A. Stern, M. Freedman, and S. Das Sarma, Rev. Mod, Phys. {\bf 80}, 1083 (2008).

\bibitem{Pachos-12}
J. K. Pachos, {\it Introduction to Topological Quantum Computation} (Canbridge University Press, New York, 2012)

\bibitem{Brouwer-11}
P. W. Brouwer, M. Duckheim, A. Romito, and F. von Oppen, Phys. Rev. Lett. {\bf 107}, 196804  (2011).

\bibitem{Lobos-12}
A. M. Lobos, R. M. Lutchyn, and S. Das Sarma, Phys. Rev. Lett. {\bf 109}, 146403 (2012).

\bibitem{DeGottardi-14}
W. DeGottardi, D. Sen, and S. Vishveshwara, Phys. Rev. Lett. {\bf 110}, 146404 (2013).

\bibitem{Altland-14}
A. Altland, D. Bagrets, L. Fritz, A. Kamenev, and H. Schmiedt, Phys. Rev. Lett. {\bf 112}, 206602  (2014).

\bibitem{Crepin-14}
F. Cr\'epin, G. Zar\'and, and P. Simon, Phys. Rev. B {\bf 90},  121407 (2014).

\bibitem{FidkowskiKitaev10}
L. Fidkowski and A. Kitaev, Phys. Rev. B {\bf 81}, 134509  (2010).

\bibitem{FidkowskiKitaev11}
L. Fidkowski and A. Kitaev, Phys. Rev. B {\bf 83}, 075103  (2011).

\bibitem{Turner11}
A. M. Turner, F. Pollmann, and E. Berg, Phys. Rev. B \textbf{83}, 075102 (2011).

\bibitem{Gurarie11}
V. Gurarie, Phys. Rev. B {\bf 83}, 085426 (2011).

\bibitem{Gangadharaiah-11}
S. Gangadharaiah, B. Braunecker, P. Simon, and D. Loss, Phys. Rev. Lett. {\bf 107}, 036801 (2011).

\bibitem{Stoudenmire-11}
E. M. Stoudenmire, J. Alicea, O. A. Starykh, and M. P.A. Fisher, Phys. Rev. B {\bf 84}, 014503 (2011).

\bibitem{Hassler-12}
F. Hassler and D. Schuricht, New J. Phys. {\bf 14}, 125018 (2012).

\bibitem{Thomale-13}
R. Thomale, S. Rachel, and P. Schmitteckert, Phys. Rev. B {\bf 88}, 161103(R)  (2013).

\bibitem{Katsura_Int_Majorana}
H. Katsura, D. Schuricht and M. Takahashi, Phys. Rev. B {\bf 92}, 115137 (2015).

\bibitem{Rahmani-15-L}
A. Rahmani, X. Zhu, M. Franz, and I. Affleck, Phys. Rev. Lett. {\bf 115}, 166401 (2015).

\bibitem{Rahmani-15-B}
A. Rahmani, X. Zhu, M. Franz, and I. Affleck, Phys. Rev. B {\bf 92}, 235123 (2015).

\bibitem{Gergs-16}
N. M. Gergs, L. Fritz, and D. Schuricht, Phys. Rev. B {\bf{93}}, 075129 (2016).

\bibitem{Grover-14}
T. Grover, D. N. Sheng, and A. Vishwanath, Science {\bf 344},  280  (2014).

\bibitem{Jian-15}
S.-K. Jian, Y.-F. Jiang, and H. Yao, Phys. Rev. Lett. {\bf 114}, 237001 (2015).

\bibitem{Hsieh-16}
T. H. Hsieh, G. B. Hal\'asz, and T. Grover, Phys. Rev. Lett. {\bf 117}, 166802 (2016).

\bibitem{Fendley12}
P. Fendley, J. Stat. Mech. (2012) P11020.

\bibitem{Clarke-14}
D. J. Clarke, J. Alicea, and K. Shtengel, Nature Phys. {\bf 10}, 877  (2014).

\bibitem{KlinovajaLoss14}
J. Klinovaja and D. Loss, Phys. Rev. Lett. {\bf 112}, 246403  (2014).

\bibitem{Mong-14}
R. S. K. Mong {\it et al.}, Phys. Rev. X {\bf 4}, 011036 (2014).

\bibitem{Jermyn-14}
A. S. Jermyn, R. S. K. Mong, J. Alicea, and P. Fendley, Phys. Rev. B {\bf 90}, 165106  (2014).

\bibitem{Aris-15}
A. Alexandradinata, N. Regnault, C. Fang, M. J. Gilbert, and B. A. Bernevig, Phys. Rev. B {\bf 94}, 125103 (2016).

\bibitem{Alicea-16}
J. Alicea and P. Fendley, Annu. Rev. Condens. Matter Phys. {\bf 7}, 119 (2016).

\bibitem{Iemini-16}
F. Iemini, C. Mora, and L. Mazza, arXiv:1611.00832.

\bibitem{Aasen-16}
D. Aasen, R. S. K. Mong, P. Fendley, J. Phys. A: Math. Theor. {\bf 49}, 354001 (2016).

\bibitem{Kitaev01}
A. {Yu Kitaev}, Phys. Usp. {\bf 44}, 131 (2001).

\bibitem{HasanKane10}
M. Z. Hasan and C. L. Kane, Rev. Mod. Phys. {\bf 82}, 3045 (2010).

\bibitem{QiZhang11}
X.-L. Qi and S.-C. Zhang, Rev. Mod. Phys. {\bf 83}, 1057 (2011)

\bibitem{shortcourse topo.ins.}
J. K. Asb\'oth, L.Oroszl\'any, and A. P\'alyi, \textit{A short Course on Topological Insulators} (Springer, 2016).

\bibitem{Kwon04}
H.-J. Kwon, K. Sengupta, and V. M. Yakovenko, Eur. Phys. J. B {\bf 37}, 349 (2004) 

\bibitem{Fu-09}
L. Fu, and C. L. Kane, Phys. Rev. B {\bf 79}, 161408(R) (2009).

\bibitem{Jiang-11}
L. Jiang, D. Pekker, J. Alicea, G. Refael, Y. Oreg, and F. von Oppen, Phys. Rev. Lett. {\bf 107}, 236401 (2011).

\bibitem{San-Jose12}
P. San-Jose, E. Prada, and R. Aguado, Phys. Rev. Lett. {\bf 108}, 257001 (2012)

\bibitem{Deng-12}
S. Deng, L. Viola, and G. Ortiz, Phys. Rev. Lett. {\bf 108}, 036803 (2012).

\bibitem{Deng-13}
S. Deng, G. Ortiz, and L. Viola, Phys. Rev. B {\bf 87}, 205414 (2013).

\bibitem{Beenakker-13-Josephson}
C. W. J. Beenakker, J. M. Edge, J. P. Dahlhaus, D. I. Pikulin,  S. Mi, and M. Wimmer, Phys. Rev. Lett. {\bf 111}, 037001 (2013).

\bibitem{Hansen-16}
E. B. Hansen, J. Danon, and K. Flensberg, Phys. Rev. B {\bf 93}, 094501 (2016).

\bibitem{Marra-16}
P. Marra, R. Citro, and A. Braggio, Phys. Rev. B {\bf 93}, 220507 (2016).

\bibitem{Alase-16}
A. Alase, E. Cobanera, G. Ortiz, and L. Viola, Phys. Rev. Lett. {\bf 117}, 076804 (2016).

\bibitem{Dmytruk-16}
O. Dmytruk, M. Trif, and P. Simon, Phys. Rev. B {\bf 94}, 115423 (2016).

\bibitem{Giuliano-16}
A. Nava, R. Giuliano, G. Campagnano, and D. Giuliano, arXiv:1612.03740. 

\bibitem{Cobanera-16}
E. Cobanera, A. Alase, G. Ortiz, and L. Viola, arXiv:1612.05567

\bibitem{def_Majorana_zero}
By Majorana zero modes in this paper, we mean a level crossing between the (many-body) ground state and the first excited state. This reduces to the presence of a zero-energy single-particle state in the non-interacting case.

\bibitem{Greiter-14}
M. Greiter, V. Schnells, and R. Thomale,  Ann. Phys. {\bf 351}, 1026 (2014). 

\bibitem{Beenakker-13}
C. W. J. Beenakker, D. I. Pikulin, T. Hyart, H. Schomerus, and J. P. Dahlhaus, Phys. Rev. Lett. {\bf 110}, 017003 (2013).

\bibitem{Ortiz-14}
G. Ortiz, J. Dukelsky, E. Cobanera, C. Esebbag, and C. W. J. Beenakker, Phys. Rev. Lett. {\bf 113}, 267002 (2014).

\bibitem{Hegde-16}
S. S. Hegde, and S. Vishveshwara, Phys. Rev. B {\bf 94}, 115166 (2016).

\bibitem{Remark_InfiniteSystem}
It is worth noting that the condition $\det M = 0$ cannot be directly applied to \textit{infinite} chains because $\det M$ cannot be simply defined in infinite chains.

\bibitem{Laplace expansion}
R. Hirota, {\it The direct method in soliton theory}, (Cambridge University Press, 2004).

\bibitem{Remark_Duality_Defect}
We here remark that the case of $\phi_{1} = \phi_{2} = \pi/2$ corresponds to the Hamiltonian limit of the Ising model with a duality defect line~\cite{Aasen-16}. The duality defect introduces the phase twist by $\pi/2$ at the position of the defect line, hence we conclude that a duality twisted Ising model has Majorana zero modes induced by the phase twist at the duality defect line, apart from the result of the existence of a domain wall in the model mentioned above.

\bibitem{Definition_Chebyshev}
The definition of the Chebyshev polynomials of the second kind $U_{n} \left( z \right)$ is
\begin{eqnarray*}
U_{n} \left( z \right) = 2 z U_{n-1} \left( z \right) - U_{n-2} \left( z \right), \\
U_{1} \left( z \right) = 2z,~U_{2} \left( z \right) = 4z^{2} - 1,
\end{eqnarray*}
and the solution of the $n$-th degree equation ``$U_{n} \left( z \right) = 0$'' is
\begin{equation*}
z = \cos \left( \frac{k}{n+1} \pi \right),~~k=1,2,\cdots,n .
\end{equation*}

\bibitem{Kao-14}
H.-C. Kao, Phys. Rev. B {\bf 90}, 245435 (2014).

\bibitem{Wakatsuki-14}
R. Wakatsuki, M. Ezawa, Y. Tanaka, and N. Nagaosa, Phys. Rev. B {\bf 90}, 014505 (2014).

\bibitem{PLA}
N. Wu, Phys. Lett. A {\bf 376}, 3530 (2012).

\bibitem{Shiozaki}
H. Shapourian, K. Shiozaki, and S. Ryu, arXiv:1607.03896.

\bibitem{Shiozaki2}
K. Shiozaki, H. Shapourian, and S. Ryu, arXiv:1609.05970.

\bibitem{Fendley-XYZ}
P. Fendley, J. Phys. A: Math. Theor. {\bf 49}, 30LT01 (2016).

\end{thebibliography}
\end{document}